\documentclass[journal]{IEEEtran}
\usepackage{gensymb}
\usepackage{graphicx}
\graphicspath{{./figure/} }
\usepackage{cite}
\usepackage{mathrsfs,amsmath}
\usepackage{epstopdf}
\usepackage{color}
\ifCLASSOPTIONcompsoc
\usepackage[caption=false,font=normalsize,labelfont=sf,textfont=sf]{subfig}
\else
\usepackage[caption=false,font=footnotesize]{subfig}
\fi

\begin{document}

\title {Signal-to-Noise Ratio of Microwave Photonic Filter With an Interferometric Structure Based on an Incoherent Broadband Optical Source}
\author{Long~Huang, \IEEEmembership{Member,~IEEE},
        Ruoming Li,
        Peng Xiang,
        Pan Dai,
        Wenxuan Wang,
        Mi Li,
        Xiangfei~Chen, \IEEEmembership{Senior~Member,~IEEE},
        and Yuechun Shi
\thanks{L. Huang, P. Dai, W. Wang, M. Li, X. Chen and Y. Shi are with Key Laboratory of Intelligent Optical Sensing and Manipulation of the Ministry of Education, and National Laboratory of Solid State Microstructures, and College of Engineering and Applied Sciences Institute of Optical Communication Engineering, Nanjing University, Nanjing 210093, China and Nanjing University (Suzhou) High-Tech Institute, Suzhou 215123, China (e-mail: shiyc@nju.edu.cn).}
\thanks{R. Li is with National Key Lab of Microwave Imaging Technology, Institute of Electronics, Chinese Academy of Sciences, Beijing, 100190, China.}
\thanks{P. Xiang is with Army Engineering University of PLA, Nanjing, 210007, China.}
\thanks{Manuscript received XXXX; accepted XXXX. Date of publication XXXX; date of current version XXXX. This work was supported by the Chinese National Key Basic Research Special Fund (2017YFA0206401); Science and Technology Project and Natural Science Foundation of Jiangsu province (BE2017003-2, BK20160907); National Natural Science Foundation of China (61435014 and 11574141); Suzhou technological innovation of key industries (SYG201844).}}

\markboth{TMTT}%
{Shell \MakeLowercase{\textit{et al.}}: Bare Demo of IEEEtran.cls for IEEE Journals}

\maketitle

\begin{abstract}
A comprehensive investigation of the signal-to-noise ratio (SNR) of the microwave photonic filter (MPF) with an interferometric structure based on an incoherent broadband optical source (IBOS) is presented from the time and frequency domains, respectively. The interferometric structure and the IBOS in the MPF result in a beneficial single-bandpass radio-frequency (RF) response. However, the IBOS adds a relatively large noise to the filtered RF signal. Therefore, the analysis of the SNR is of great importance for the design of this kind of MPFs. Theoretical analysis shows that the SNR is a function of the center frequency of the passband, the modulation index, the chromatic dispersion, and the shape of the IBOS. An experiment is performed to verify the theory, and experimental results agree well with the theoretical calculation.
\end{abstract}


\section{Introduction}
\IEEEPARstart{M}{icrowave} photonic filters (MPFs) have attracted much research interest in radio frequency (RF) signal processing as the electronic bottleneck can be circumvented with the advantages provided by photonic devices such as large time-bandwidth product, high tunability, high reconfigurability, and electromagnetic interference (EMI) immunity \cite{J. Yao, Capmany1}. Generally, MPFs can be implemented with infinite impulse response (IIR) or finite impulse response (FIR) structures \cite{Capmany1}. The discrete taps in the IIR or the FIR structure can be produced by a laser array \cite{Capmany_mtt} or an optical frequency comb \cite{Hamidi_mtt}. As a low-cost alternative of the laser array and the optical frequency comb, the sliced incoherent broadband optical source (IBOS) has been proposed to generate the discrete taps. Optical filters (OFs) such as fiber Bragg gratings \cite{Hunter,Mora1,Mora2,Popov}, Fabry-Perot filters \cite{Capmany2} and array waveguide gratings \cite{Pastor} can be used to slice the IBOS. However, the passbands of the MPF with the IIR or the FIR structure are periodic due to the discrete taps. Therefore, the processing frequency of the MPF is restricted to the free spectral range (FSR) of the periodic passbands in order to avoid spectral overlapping. Furthermore, if the center frequency of the MPF with discrete taps is tuned, the shape of the passbands will be changed \cite{J. Yao, Capmany1}.
On the other hand, the IBOS combined with an interferometric structure can result in a beneficial single-bandpass RF response. By adjusting the time delay of the interferometric structure, the center frequency of the passband can be tuned without changing the shape.
\par Due to the merits of the single-bandpass response, the invariant passband shape, and the cost-effective setup, many single-bandpass MPFs based on the IBOS have been proposed. These MPFs can be classified in terms of their employed modulators, such as the Mach-Zehnder modulator (MZM) \cite{Mora3}, the single-sideband (SSB) modulator \cite{Xu}, the phase modulator \cite{Xue}, the polarization modulator \cite{Wang}, and the dual-input modulator \cite{Li}. Moreover, the IBOS-based single-bandpass MPF has been widely used in radio-over-fiber (RoF) communication systems \cite{P_Li, Chi}, optoelectronic oscillators (OEOs) \cite{C_Li, Zhang} and microwave frequency multiplication \cite{Gao}. However, due to the incoherence of the IBOS, an excess intensity noise is added to the filtered RF signal after optical-to-electrical conversion. In the OEO, the excess intensity noise is converted to the phase noise of the oscillating microwave signal. In the RoF system, the signal-to-noise ratio (SNR) sets the limit of the channel capacity. Therefore, the analysis of the SNR is of great importance for the design of the IBOS-based single-bandpass MPF. Previously, the SNR of the digital optical communication system based on the IBOS is investigated in \cite{Pendock_JLT,Pendock_PTL,Lee}, the SNR of the optical beamformer based on the IBOS is investigated in \cite{Huang_JQE}, and the SNR of the IBOS-based MPF with discrete taps is investigated in \cite{Yi,F1,F2}. Nevertheless, although many MPFs with an interferometric structure based on the IBOS have been proposed \cite{Mora3, Xu, Xue, Wang, Li}, the SNR of the output RF signal has not been measured, let alone theoretically analyzed.
\par In this paper, we give a comprehensive analysis of the SNR of the IBOS-based MPF with an interferometric structure for the first time. More specifically, the SNR of the filtered RF signal at the center frequency of the passband is investigated. Firstly, the previous schemes are reviewed. Subsequently, we introduce a general configuration which can represent the IBOS-based MPFs with an interferometric structure reported in the literature. Based on the general configuration, we present a theoretical analysis of the SNR. In the analysis, the signal and the noise power spectrum densities (PSDs) are derived, respectively, and then the SNR which is the ratio of the signal and noise PSDs is obtained. The signal PSD is actually the magnitude response of the MPF, and the SNR is a function of the center frequency of the passband, the modulation index, the chromatic dispersion and the shape of the IBOS. We perform an experiment to verify the analysis. The measured SNR as well as the magnitude response show good agreement with the theoretical calculation.
\par Notation: $\Re$ represents the real part of a complex number. $\left \langle {\cdot} \right \rangle$ represents ensemble average. $\overline{\cdot}$ denotes time average. $\otimes$ denotes convolution. $\mathscr{F}[.]$ denotes the Fourier transform.
\section{General system setup}
In this section, previous schemes of the MPF with an interferometric structure based on an IBOS \cite{Mora3, Xu, Xue, Wang, Li} are reviewed. Then we introduce a general configuration which includes the previous configurations as special cases. Based on this general configuration, an analysis framework for the SNR is presented.
\begin{figure}[htbp]
\centering
\includegraphics [width=3.2in] {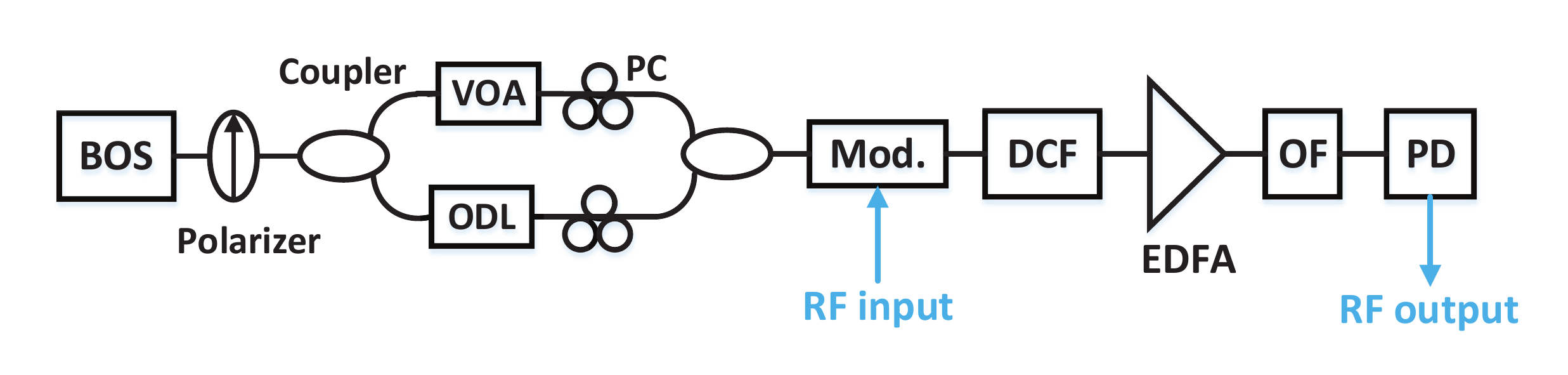}
\caption{Setup of the IBOS-based single-bandpass MPF where delayed and non-delayed arms are modulated via a same modulator.}
\label{setup1}
\end{figure}
\par We suppose the interferometric structure in the MPF has two arms, and the time delay of one of the arms can be tuned. As the first typical configuration of the MPF, the delayed and non-delayed arms are modulated via a same modulator. The schematic diagram of the setup is shown in Fig. \ref{setup1} \cite{Mora3, Xu}. The IBOS is generated by the amplified spontaneous emission (ASE) of an Erbium doped fiber amplifier (EDFA). A polarizer is placed after the IBOS to coerce the light to be linearly polarized. The light after the polarizer is then split by a 3-dB optical coupler into two arms. One arm is connected to a variable optical attenuator (VOA) followed by a polarization controller (PC) while the other arm is connected to an optical delay line (ODL) followed by another PC. The VOA is adjusted to balance the power of the two arms, and the ODL is adjusted to tune the relative time delay between the two arms. Afterwards, the split lights are combined by another 3-dB coupler whose output is connected to an optical modulator. The PC in each arm is adjusted to align the polarization state of the light to the modulator. A dispersion compensation fiber (DCF) module is placed after the modulator to introduce chromatic dispersion to the light. An EDFA is connected to the output of the DCF to amplify the light. A tunable OF is incorporated after the EDFA to tailor the shape of the light. Finally, the light is sent to a photodetector (PD) for optical-to-electrical conversion. The RF signal is applied to the optical modulator, and the filtered RF signal is obtained at the output of the PD.
\par In the MPF, the broadband light generated by the IBOS and filtered by the OF is a bandpass signal which can be expressed as ${\varepsilon _0}(t) = A(t)\cos \left[ {2\pi {f_0}t + \theta (t)} \right]$. Its analytic signal representation can be given by
\begin{equation}
{E_a}(t) = A(t){e^{j\left[ {2\pi {f_0}t + \theta (t)} \right]}} = {E_0}(t){e^{j2\pi {f_0}t}},
\end{equation}
where $E_0(t)$ and $f_0$ are the complex envelope and the center frequency of the light, respectively. Mathematically, $E_a(t)$ and $E_0(t)$ are stationary complex circular Gaussian stochastic processes \cite{Goodman}. Stationarity implies that the autocorrelation function only depends on the lag variable and has Hermitian symmetry ${R}(u) = R^*( - u)$. The autocorrelation functions of ${E_a}(t)$ and ${E_0}(t)$ are denoted by ${R_a}(u)$ and ${R_0}(u)$, respectively, and then the relationship between them can be given by
\begin{IEEEeqnarray*}{rCl}
{R_a}(u) &=& \left\langle {E_a^*(t){E_a}(t + u)} \right\rangle \\
 &=& \left\langle {E_0^*(t){E_0}(t + u)} \right\rangle {e^{j2\pi {f_0}u}} = {R_0}(u){e^{j2\pi {f_0}u}}.
\IEEEyesnumber
\end{IEEEeqnarray*}
Based on the Wiener-Khinchin theorem, the power spectral density (PSD) of ${E_a}(t)$ and ${E_0}(t)$ is the Fourier transform of the autocorrelation function, i.e. ${G_a}(f) = \mathscr{F}\left[ {{R_a}(u)} \right]$ and $G(f) = \mathscr{F}\left[ {{R_0}(u)} \right]$. According to (2), we have  ${G_a}(f) = G(f - {f_0})$. Generally, ${G_a}(f)$ centered at $f_0$ can be measured by an optical spectrum analyzer (OSA), and $G(f)$ at baseband is more convenient for calculation.
\par In the MPF shown in Fig. \ref{setup1}, the light before the DCF can be expressed as
\begin{IEEEeqnarray*}{rCl}
{\varepsilon _D}(t) &\propto& \Re \left\{ \left[{E_0}(t){e^{j2\pi {f_0}t}} + {E_0}(t - d){e^{j2\pi {f_0}(t - d)}} \right] m(t)\right\}\\
&=& \Re \left\{ {E_D}(t){e^{j2\pi {f_0}t}} \right\},\IEEEyesnumber
\end{IEEEeqnarray*}
where
\begin{equation}\label{eq:mzm}
{E_D}(t) = {E_0}(t)m(t) + {E_0}(t - d){e^{ - j2\pi {f_0}d}}m(t)\\
\end{equation}
is the envelope of the light, $d$ is the time delay introduced by the ODL, and $m(t)$ is the modulation function.
\begin{figure}[htbp]
\centering
\includegraphics [width=3.2in] {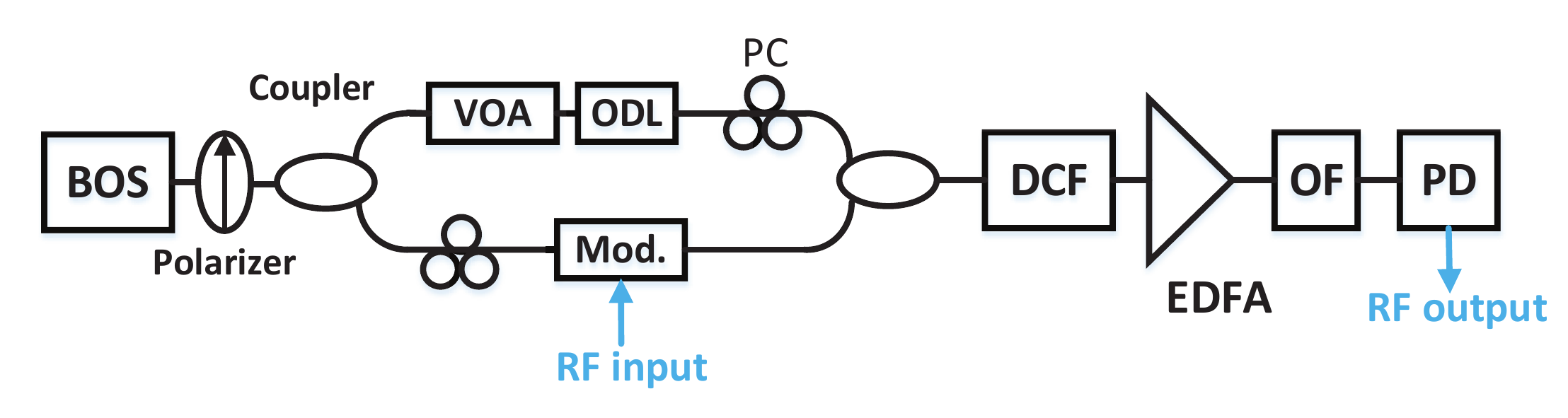}
\caption{Setup of the IBOS-based single-bandpass MPF where one arm is not modulated.}
\label{setup2}
\end{figure}
\par Another typical configuration of the MPF is illustrated in Fig. \ref{setup2} \cite{Xue}. In this configuration, only one of the two arms is modulated by a RF signal. The IBOS is polarized and split by a 3-dB coupler. One arm is connected to a VOA followed by an ODL and a PC, while the other arm is connected to a modulator through a PC. The two arms are then combined and sent to the DCF. After the DCF, the EDFA and the OF, the light is finally connected to a PD for optical-to-electrical conversion. According to \cite{Xue}, the light before the DCF can be expressed as
\begin{IEEEeqnarray*}{rCl}
{\varepsilon _D}(t) &=& \Re \left\{ {E_0}(t - d){e^{j2\pi {f_0}(t - d)}} \!+\! {E_0}(t){e^{j2\pi {f_0}t}}m(t) \right\}. \IEEEyesnumber
\end{IEEEeqnarray*}
Therefore, the envelope of the light can be expressed as
\begin{equation} \label{eq:pm}
{E_D}(t) = {E_0}(t - d){e^{ - j2\pi {f_0}d}} + {E_0}(t)m(t).
\end{equation}
\begin{figure}[htbp]
\centering
\includegraphics [width=3.2in] {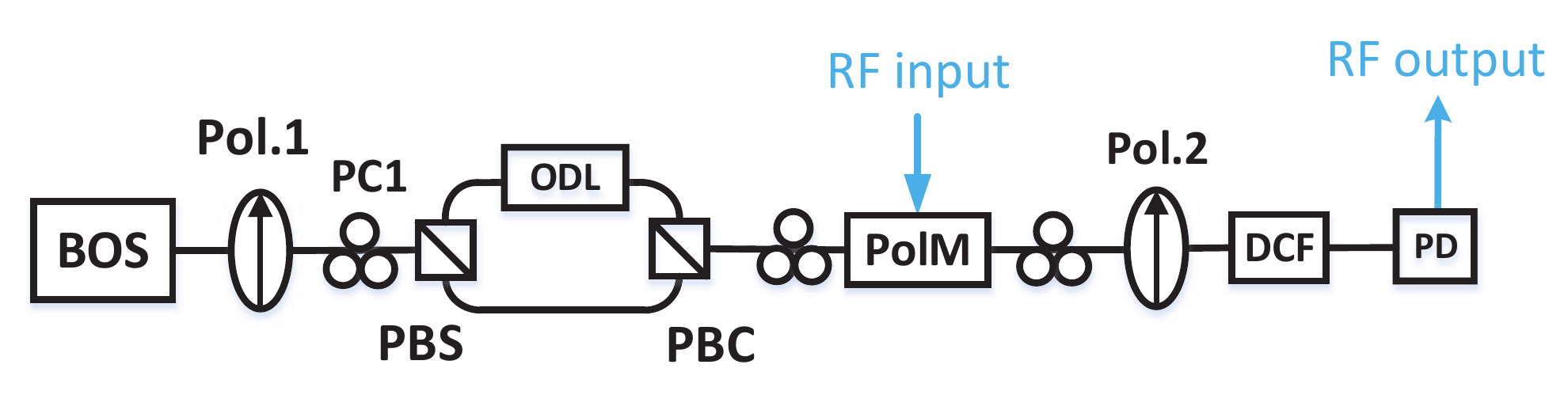}
\caption{Setup of the IBOS-based single-bandpass MPF using a polarization modulator (PolM).}
\label{polm}
\end{figure}
\par In \cite{Wang}, a single-bandpass MPF using a polarization modulator is proposed, and the schematic diagram of the configuration is shown in Fig. \ref{polm}. The light before the DCF is given by
\begin{equation}
{\varepsilon _{D}}(t) = \Re \left\{ {J_0}(\gamma){\varepsilon _0}(t-d) \!+\! j2{J_1}(\gamma)\cos(2\pi {f_m}t){\varepsilon _0}(t)\right\}\!\!.
\end{equation}
Hence the envelope before the DCF can be given by
\begin{equation}\label{eq:polm}
{E_{D}}(t) = k {E_0}(t-d)e^{ - j2\pi {f_0}d} + {E_0}(t)m(t),
\end{equation}
where $k = {J_0}(\gamma )$ and $m(t) = j2{J_1}(\gamma )\cos (2\pi {f_m}t)$. From (\ref{eq:pm}) and (\ref{eq:polm}), we can see the configuration shown in Fig. 3 is actually equivalent to that shown in Fig. 2.
\begin{figure}[htbp]
\centering
\includegraphics [width=3.1in] {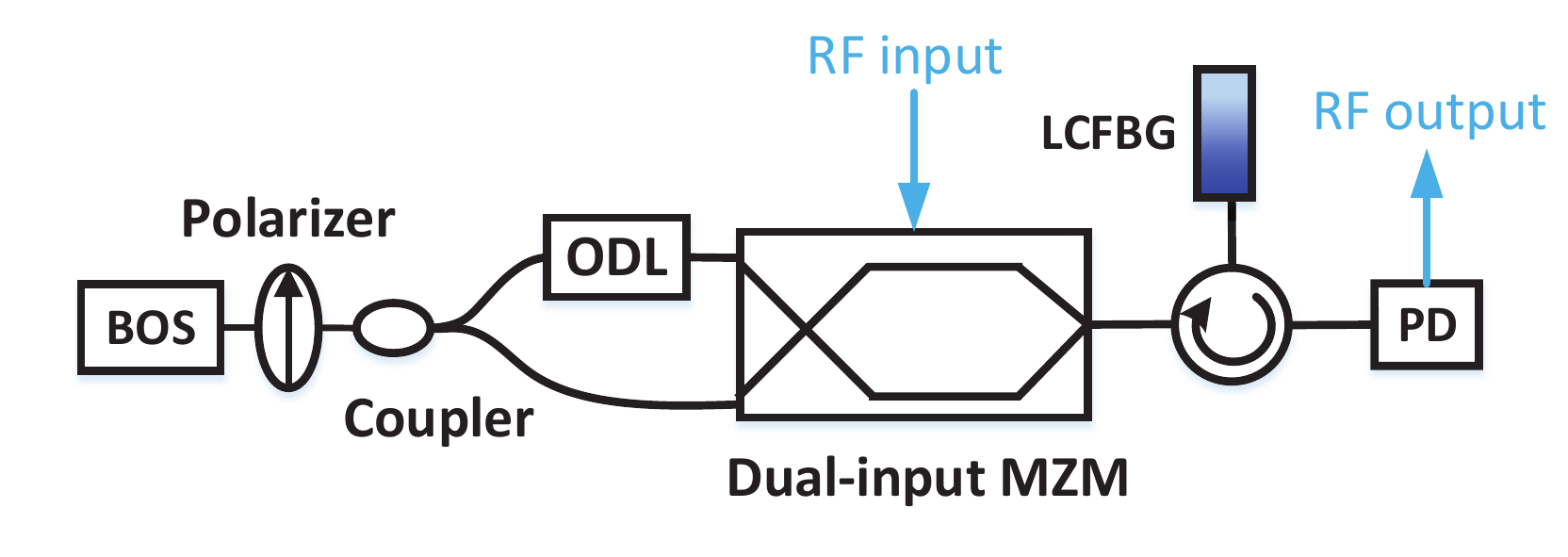}
\caption{Setup of the IBOS-based single-bandpass MPF using a dual-input MZM.}
\label{dualmzm}
\end{figure}

\par In \cite{Li}, another IBOS-based single-bandpass MPF employing a dual-input MZM is demonstrated. The schematic diagram of the configuration is shown in Fig. \ref{dualmzm}. The linearly chirped fiber Bragg grating (LCFBG) is used to introduce chromatic dispersion to the light. According to \cite{Li}, the light before the LCFBG can be expressed as
\begin{IEEEeqnarray*} {rCl}
{\varepsilon _D}(t) &=& \Re \big\{ j\sin \left[ {\pi {V_{bias}}/{V_\pi } + m\cos (2{\pi}{f_m}t)} \right]{\varepsilon _0}(t - d)\\
 &&- \cos \left[ {\pi {V_{bias}}/{V_\pi } + m\cos (2{\pi}{f_m}t)} \right]{\varepsilon _0}(t) \big\}, \IEEEyesnumber
\end{IEEEeqnarray*}
where $V_{bias}$ is the bias voltage applied to the dual-input MZM, $V_\pi$ is half-wave voltage of the dual-input MZM, and $m$ is the modulation index. In
\cite{Li}, the dual-input MZM is biased at the quadrature point, so the envelope of the light before the DCF can be given by
\begin{equation}\label{eq:dimzm}
{E_D}(t) = k{E_0}(t - d){e^{ - j2\pi {f_0}d}} + {E_0}(t)m(t),
\end{equation}
where $k = j{J_0}(m)$ and $m(t) = 2{J_1}(m)\cos ({\omega _m}t)$. We can see this scheme is also equivalent to the scheme shown in Fig. \ref{setup2}.
\par
\begin{figure}[htbp]
\centering
\includegraphics [width=3.4in] {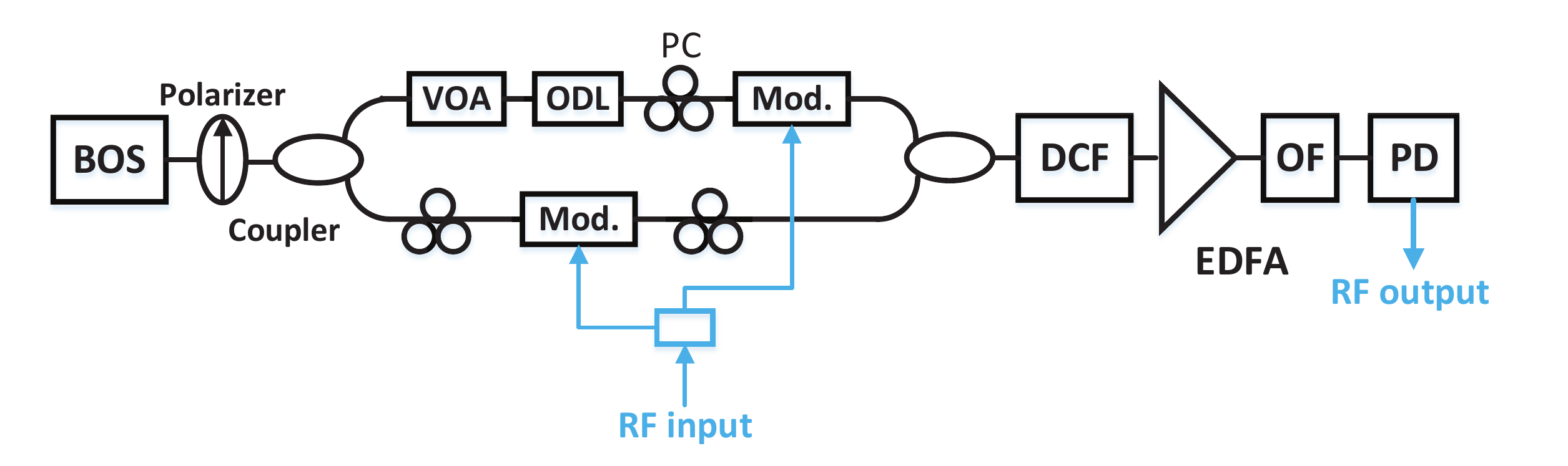}
\caption{General configuration of the IBOS-based single-bandpass MPF.}
\label{gen_mod}
\end{figure}

 Based on the review of the previous schemes, we propose a general configuration which is shown in Fig. {\ref{gen_mod}}. In this general configuration, the two arms are modulated via different modulators. Therefore, the configuration shown in Fig. \ref{setup1} is equivalent to the general configuration in which the two modulators are identical. The configurations shown in Figs. \ref{setup2}, \ref{polm}, \ref{dualmzm} are equivalent to the proposed general configuration in which only one arm is modulated. As a conclusion, the existing IBOS-based single-bandpass MPFs in \cite{Mora3, Xu, Xue, Wang, Li} can all be regarded as special cases of the proposed MPF model shown in Fig. \ref{gen_mod}. Using the general configuration, we present a framework to analyze the SNR.

In the general configuration, the envelope of the light before the dispersive medium can be given by
\begin{equation}
{E_D}(t) = {E_0}(t){m_1}(t) + {E_0}(t - d){e^{ - j2\pi {f_0}d}}{m_2}(t),
\end{equation}
where $m_1(t)$ and $m_2(t)$ are two different modulation functions. The field envelope after propagation in a dispersive medium is given by solving the optical field envelope equation \cite{Marshall}, i.e.,
\begin{equation} \label{eq:dispersive}
\frac{{\partial {E_D(t)}}}{{\partial z}} + {\beta _0}^\prime \frac{{\partial {E_D(t)}}}{{\partial t}} - \frac{j}{2}{\beta _0}^{\prime \prime }\frac{{{\partial ^2}{E_D(t)}}}{{{\partial ^2}t}} = 0,
\end{equation}
where ${\beta '_0}$ is the group delay per unit length, and ${\beta ''_0}$ is the group delay dispersion per unit angular frequency per unit length. It is proved that the double-sided intensity PSD of the light after the dispersive medium can be expressed as \cite{F1} \cite{Marshall}
\begin{IEEEeqnarray*} {rCl}
{S}(f) &=& \mathscr{F} \left[ {{R_{4{E_D}}}}(v,v+u,u) \right] \\
&=& \int_{ - \infty }^{ + \infty } {{R_{4{E_D}}}(v,v + u,u){e^{ - j2\pi fu}}du}, \IEEEyesnumber \label{eq:S(f)}
\end{IEEEeqnarray*}
where
\begin{equation}
v = 2\pi f\phi
\end{equation}
is an auxiliary variable used to simplify the notation, $\phi  = {\beta ''_0}z$ is the overall group delay dispersion with length $z$, and ${R_{4E_D}} (v,v+u,u)$ represents the time average of fourth-order moment of $E_D(t)$, which is given by
\begin{IEEEeqnarray*}{l}
\IEEEeqnarraymulticol{1}{l}{
{R_{4{E_D}}}(v,v + u,u)}\\
= \overline {\left\langle {E_D^{\rm{*}}(t){E_D}(t + v)E_D^*(t + v + u){E_D}(t + u)} \right\rangle}\\
= \left[ {{{\left| {{R_0}(v)} \right|}^2} + {{\left| {{R_0}(u)} \right|}^2}} \right]\\
\quad \cdot \overline{ m_1^*(t){m_1}(t + v)m_1^*(t + v + u){m_1}(t + u)}\\
 + \left[ {{R_0}(v)R_0^*(v + d) + {R_0}(u - d)R_0^*(u)} \right]{e^{ - j2\pi{f_0}d}}\\
 \quad \cdot \overline {m_1^*(t){m_1}(t + v)m_1^*(t + v + u){m_2}(t + u)}\\
 + \left[ {{R_0}(v)R_0^*(v - d) + {R_0}(u)R_0^*(u - d)} \right]{e^{j2\pi{f_0}d}}\\
 \quad \cdot \overline {m_1^*(t){m_1}(t + v)m_2^*(t + v + u){m_1}(t + u)}\\
 + \left[ {{{\left| {{R_0}(v)} \right|}^2} + {{\left| {{R_0}(u - d)} \right|}^2}} \right]\\
 \quad \cdot \overline {m_1^*(t){m_1}(t + v)m_2^*(t + v + u){m_2}(t + u)}\\
 + \left[ {{R_0}(v - d)R_0^*(v) + {R_0}(u)R_0^*(u + d)} \right]{e^{ - j2\pi{f_0}d}}\\
 \quad \cdot \overline {m_1^*(t){m_2}(t + v)m_1^*(t + v + u){m_1}(t + u)}\\
 + \left[ {{R_0}(v - d)R_0^*(v + d) + {R_0}(u - d)R_0^*(u + d)} \right]{e^{ - j4\pi{f_0}d}}\\
 \quad \cdot \overline {m_1^*(t){m_2}(t + v)m_1^*(t + v + u){m_2}(t + u)}\\
 + \left[ {{{\left| {{R_0}(v-d)} \right|}^2} + {{\left| {{R_0}(u)} \right|}^2}} \right]\\
 \quad \cdot \overline {m_1^*(t){m_2}(t + v)m_2^*(t + v + u){m_1}(t + u)}\\
 + \left[ {{R_0}(v - d)R_0^*(v) + {R_0}(u - d)R_0^*(u)} \right]{e^{ - j2\pi{f_0}d}}\\
 \quad \cdot \overline {m_1^*(t){m_2}(t + v)m_2^*(t + v + u){m_2}(t + u)}\\
 + \left[ {{R_0}(v + d)R_0^*(v) + {R_0}(u + d)R_0^*(u)} \right]{e^{j2\pi{f_0}d}}\\
 \quad \cdot \overline {m_2^*(t){m_1}(t + v)m_1^*(t + v + u){m_1}(t + u)}\\
 + \left[ {{{\left| {{R_0}(v + d)} \right|}^2} + {{\left| {{R_0}(u)} \right|}^2}} \right]\\
 \quad \cdot \overline {m_2^*(t){m_1}(t + v)m_1^*(t + v + u){m_2}(t + u)}\\
 + \left[ {{R_0}(v + d)R_0^*(v - d) + {R_0}(u + d)R_0^*(u - d)} \right]{e^{j4\pi{f_0}d}}\\
 \quad \cdot \overline {m_2^*(t){m_1}(t + v)m_2^*(t + v + u){m_1}(t + u)}\\
 + \left[ {{R_0}(v + d)R_0^*(v) + {R_0}(u)R_0^*(u - d)} \right]{e^{j2\pi{f_0}d}}\\
 \quad \cdot \overline {m_2^*(t){m_1}(t + v)m_2^*(t + v + u){m_2}(t + u)}\\
 + \left[ {{{\left| {{R_0}(v)} \right|}^2} + {{\left| {{R_0}(u + d)} \right|}^2}} \right]\\
 \quad \cdot \overline {m_2^*(t){m_2}(t + v)m_1^*(t + v + u){m_1}(t + u)}\\
 + \left[ {{R_0}(v)R_0^*(v + d) + {R_0}(u)R_0^*(u + d)} \right]{e^{ - j2\pi{f_0}d}}\\
 \quad \cdot \overline {m_2^*(t){m_2}(t + v)m_1^*(t + v + u){m_2}(t + u)}\\
 + \left[ {{R_0}(v)R_0^*(v - d) + {R_0}(u + d)R_0^*(u)} \right]{e^{j2\pi{f_0}d}}\\
 \quad \cdot \overline {m_2^*(t){m_2}(t + v)m_2^*(t + v + u){m_1}(t + u)}\\
 + \left[ {{{\left| {{R_0}(v)} \right|}^2} + {{\left| {{R_0}(u)} \right|}^2}} \right]\\
 \quad \cdot \overline {m_2^*(t){m_2}(t + v)m_2^*(t + v + u){m_2}(t + u)}.
 \IEEEyesnumber \label{eq:general_autocor}
\end{IEEEeqnarray*}
In deriving (\ref{eq:general_autocor}), we use Gaussian moment theorem \cite{Goodman} which states if $g_1$, $g_2$, $g_3$ and $g_4$ are circular complex Gaussian random variables, then $\left\langle {g_1^*{g_2}g_3^*{g_4}} \right\rangle  = \left\langle {g_1^*{g_2}} \right\rangle \left\langle {g_3^*{g_4}} \right\rangle  + \left\langle {g_1^*{g_3}} \right\rangle \left\langle {g_2^*{g_4}} \right\rangle$.
\par For a specific setup of the general configuration, specific $m_1(t)$ and $m_2(t)$ can be substituted into (\ref{eq:general_autocor}). Then the PSD of the MPF can be derived by utilizing (\ref{eq:S(f)}). In Section III, the SNRs of two specific MPFs are derived by the framework.

\section{Signal-to-Noise analysis}
In this section, the SNR of two typical configurations of the IBOS-based MPF with an interferometric structure is presented. We choose the two configurations since we can perform an experiment to verify the theoretical analysis in our lab. The experiment results which verify the theoretical analysis are given in Section IV.
\subsection{The MPF in which two arms are modulated via a same modulator}

The schematic diagram of the MPF in which two arms are modulated via a same modulator is shown in Fig. 1 as the first typical configuration. In this configuration, $m_1(t)=m_2(t)=m(t)$. Then (\ref{eq:general_autocor}) can be simplified to
\begin{equation}
{R_{4{E_D}}}(v,v + u,u) = \left[ {\left| {H(v)} \right|^2} + {\left| {H(u)} \right|^2} \right]{R_{4m}}(v,v + u,u),
\end{equation}
where
\begin{equation} \label{eq:H}
H(x) = 2{R_0}(x) + {R_0}(x - d){e^{ - j2\pi {f_0}d}} + {R_0}(x + d){e^{j2\pi {f_0}d}},
\end{equation}
where $x$ = $v$ or $u$, and $R_{4m}$ is the time average of fourth-order moment of $m(t)$.
\par Next, we calculate $R_{4m}(v,v+u,u)$. We suppose a pure RF signal with a frequency $f_m$ is applied to the modulator. Therefore, $m(t)$ is a periodic function which can be expanded by Fourier series, i.e.,
\begin{equation}
m(t) = \sum\limits_n {{M_n}{e^{jn2\pi {f_m}t}}}.
\end{equation}
${m^*}(t)m(t + v)$ can be also expanded by Fourier series,
\begin{equation}
{m^*}(t)m(t + v) = \sum\limits_s {R_{2m}^{(s)}(v){e^{j2\pi {f_m}st}}},
\end{equation}
where ${R_{2m}^{(s)}(v)}$ are s-th cyclic autocorrelation functions of $m(t)$ \cite{F2}, and their expressions can be given by
\begin{IEEEeqnarray*}{rCl}
R_{2m}^{(s)}(v) &=& \overline {{m^*}(t)m(t + v){e^{ - j2\pi {f_m}st}}} \\
 &=& \overline {\sum\limits_p {M_p^*{e^{ - j2\pi {f_m}pt}}\sum\limits_q {{M_q}{e^{jmq(t + v)}}{e^{ - j2\pi {f_m}st}}} } } \\
 &=& \sum\limits_q {{M_q}M_{q - s}^*} {e^{j2\pi {f_m}qv}}. \IEEEyesnumber
\end{IEEEeqnarray*}
As a result, ${R_{4m}}(v,v+u,u)$ can be expressed in terms of s-th cyclic autocorrelation functions, i.e.,
\begin{IEEEeqnarray*}{cRl}
\IEEEeqnarraymulticol{3}{l}{
{R_{4m}}(v,v + u,u)}\\
&=& \overline { \big\langle {m^*}(t)m(t + v){m^*}(t + v + u)m(t + u)\big\rangle } \\
&=& \overline {\sum\limits_s {R_{2m}^{(s)}(v){e^{j2\pi {f_m}st}}}\sum\limits_r {R_{2m}^{(r)*}(v){e^{ - j2\pi {f_m}r(t + u)}}}}\\
&=& {\sum\limits_s {\left| {R_{2m}^{(s)}(v)} \right|} ^2}{e^{ - js2\pi {f_m}u}}. \IEEEyesnumber \label{R4m}
 \end{IEEEeqnarray*}
In our analysis, small-signal modulation is assumed. If the employed modulator is an MZM which synthesizes double-sideband modulation (DSB) \cite{Mora3} signals, $m(t)$ can be expressed as
\begin{equation}
m(t) \approx 1 + (\gamma/2) {e^{j2\pi {f_m}t}} + (\gamma/2) {e^{ - j2\pi {f_m}t}}.
\end{equation}
 Then, s-th cyclic autocorrelation functions for DSB modulation are calculated as $R_{2m}^{(0)}(v) = 1 + ({\gamma ^2}/2){e^{j2\pi {f_m}v}}$ and $R_{2m}^{(1)}(v) = R_{2m}^{( - 1)}(v) = (\gamma /2)\left[ {{e^{j2\pi {f_m}v}} + 1} \right]$.
If the modulator is a single-sideband (SSB) modulator \cite{Xu}, $m(t)$ can be expressed as
\begin{equation}
m(t) \approx 1 + (\gamma/2) {e^{j2\pi {f_m}t}}.
\end{equation}
Hence s-th cyclic autocorrelation functions for SSB modulation are calculated as $R_{2m}^{(0)}(v) = 1 + {(\gamma/2) ^2}{e^{j2\pi {f_m}v}},$
$R_{2m}^{(1)}(v) = (\gamma/2) {e^{j2\pi {f_m}v}},$ and $R_{2m}^{( - 1)}(v) = \gamma/2.$
\par Utilizing (\ref{R4m}), the expression of $R_{4E_D}(v,v+u,u)$ can be given by
\begin{IEEEeqnarray*} {rcl}
\IEEEeqnarraymulticol{3}{l}{
{R_{4{E_D}}}(v,v + u,u)}\\
 &=& \left[ {{{\left| {H(v)} \right|}^2} + {{\left| {H(u)} \right|}^2}} \right]\cdot\\
&&\bigg[ {{{\left| {R_{2m}^{(0)}(v)} \right|}^2} + {{\left| {R_{2m}^{(1)}(v)} \right|}^2}{e^{-j2\pi {f_m}u}} + {{\left| {R_{2m}^{( - 1)}(v)} \right|}^2}{e^{j2\pi {f_m}u}}} \bigg]. \\
\IEEEyesnumber
\end{IEEEeqnarray*}
We apply Fourier transform to $R_{4E_D}(v,v+u,u)$ with respect to $u$, and then the intensity PSD can be obtained as
\begin{IEEEeqnarray*} {rCl}
\IEEEeqnarraymulticol{3}{l}{
S(f) = \mathscr{F} \left[ {{R_{4{E_D}}}}(v,v+u,u) \right] }\\
 &=& \bigg[ {\left| {H(0)} \right|^2}{\left| {R_{2m}^{(0)}(0)} \right|^2}\delta (f) \\
 &&+ {\left| {H( - {v_m})} \right|^2}{\left| {R_{2m}^{(1)}( - {v_m})} \right|^2}\delta (f + {f_m})\\
 &&+ {\left| {H({v_m})} \right|^2}{\left| {R_{2m}^{( - 1)}({v_m})} \right|^2}\delta (f - {f_m}) \bigg]\\
 &&+ \bigg[ {\left| {R_{2m}^{(0)}(v)} \right|^2}{S_H}(f)\\
 &&+ {\left| {R_{2m}^{(1)}(v)} \right|^2}{S_H}(f + {f_m}) + {\left| {R_{2m}^{( - 1)}(v)} \right|^2}{S_H}(f - {f_m}) \bigg],\\ \IEEEyesnumber \label{eq:S_with}
\end{IEEEeqnarray*}
where $S_H(f)= \mathscr{F} \left[ {{{\left| {H(u)} \right|}^2}} \right] $. The signal and noise PSD can be respectively given by
\begin{IEEEeqnarray*} {rCl}
S^{sig}(f) &=& \bigg[ {\left| {H(0)} \right|^2}{\left| {R_{2m}^{(0)}(0)} \right|^2}\delta (f) \\
 &&+ {\left| {H( - {v_m})} \right|^2}{\left| {R_{2m}^{(1)}( - {v_m})} \right|^2}\delta (f + {f_m})\\
 &&+ {\left| {H({v_m})} \right|^2}{\left| {R_{2m}^{( - 1)}({v_m})} \right|^2}\delta (f - {f_m}) \bigg].
 \IEEEyesnumber \label{eq:sig}
\end{IEEEeqnarray*}
\begin{IEEEeqnarray*} {rCl}
S^{no}(f) &=& \bigg[ {\left| {R_{2m}^{(0)}(v)} \right|^2}{S_H}(f)+ {\left| {R_{2m}^{(1)}(v)} \right|^2}{S_H}(f + {f_m}) \\
&&+ {\left| {R_{2m}^{( - 1)}(v)} \right|^2}{S_H}(f - {f_m}) \bigg].
\IEEEyesnumber \label{eq:no}
\end{IEEEeqnarray*}

\par
\begin{figure}[htbp]
\centering
\includegraphics [width=3.1in] {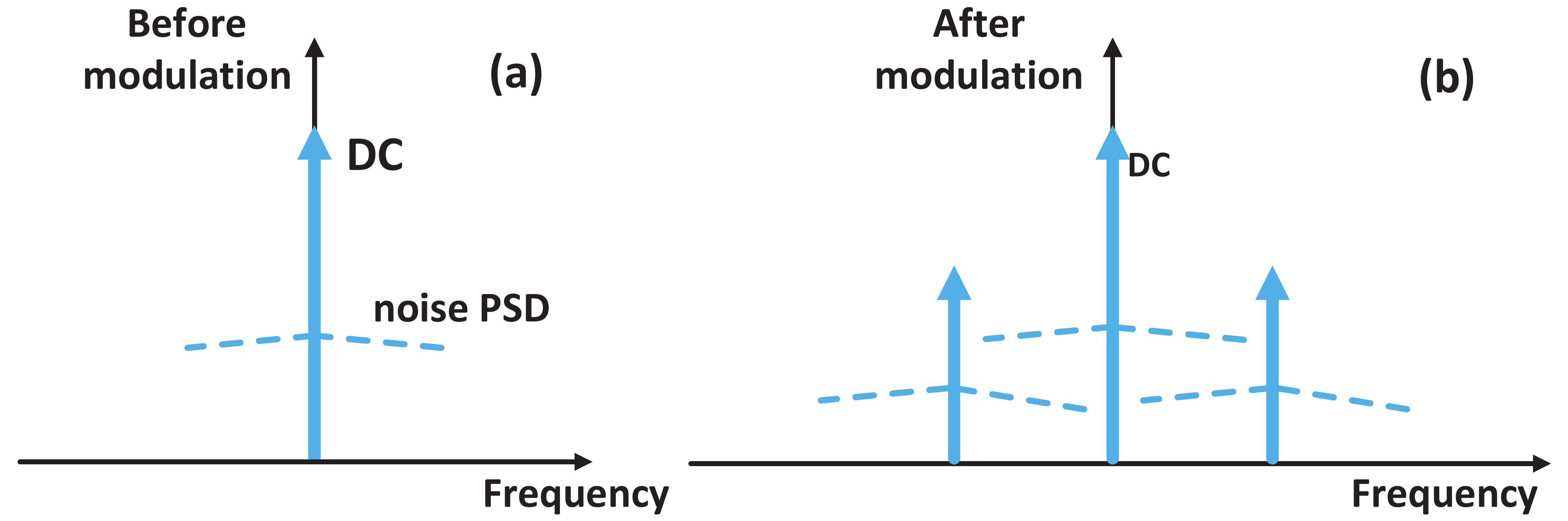}
\caption{Intensity PSD before and after modulation.}
\label{psd}
\end{figure}

 If there is no modulation, i.e $\gamma=0$, then the intensity PSD can be given by
\begin{equation} \label{eq:S_wo}
S_{wo}(f) = {\left| {H(0)} \right|^2}\delta (f) + {S_H}(f).
\end{equation}
According to (\ref{eq:S_wo}) and (\ref{eq:S_with}), the PSD before and after modulation is shown in Fig. \ref{psd}. As can be seen, after applying a RF signal, the baseband noise signal is up-converted to the RF.
\par For the DSB modulation, by using the sifting property of the delta function ($\int_{ - \infty }^{ + \infty } {f(x)\delta (x - a)dx = f(a)} $), we can obtain the power of the signal by integrating $S^{sig}(f)$ at $\pm f_m$,
\begin{IEEEeqnarray*} {l}
{P_{dsb}^{sig}}({f_m}) = 2{(\gamma/2) ^2}{\cos ^2}\left( {\pi {f_m}{v_m}} \right){\left| {H({v_m})} \right|^2}.
 \IEEEyesnumber \label{eq:dsbdsb}
\end{IEEEeqnarray*}
As can be seen from (\ref{eq:dsbdsb}), the frequency response is ${\left| {H\left( {{v_m}} \right)} \right|^2}$, which is shaped by the term ${\cos ^2}\left( {\pi {f_m}{v_m}} \right)$ originated from the chromatic dispersion induced power fading.
For the SSB modulation, the power of the signal can be obtained in the same manner,
\begin{IEEEeqnarray*} {l}
{P_{ssb}^{sig}}({f_m}) = 2{(\gamma/2) ^2}{\left| {H({v_m})} \right|^2}.
\IEEEyesnumber \label{eq:P_ssb}
\end{IEEEeqnarray*}
As can be seen from (\ref{eq:P_ssb}), the chromatic dispersion induced power fading effect is eliminated, so we only investigate the SNR of the SSB modulation.
If the shape of the OF is rectangular, then the baseband PSD can be given by
\begin{equation} \label{eq:Gf}
G(f) = \left\{
\begin{IEEEeqnarraybox} [] [c] {l?s}
\IEEEstrut
N_{0} & $\left| f \right| \le B/2$, \\
0 & otherwise.
\IEEEstrut
\end{IEEEeqnarraybox}
\right.
\end{equation}
where $N_0$ is the magnitude and $B$ is the optical bandwidth.
Therefore, the autocorrelation function $R_0(v)$ can be expressed as
\begin{equation}\label{eq:R0}
{R_0}({v_m}) = {\left. {{\mathscr{F}^{ - 1}}\left[ {G(f)} \right]} \right|_{{v_m}}} = {N_0}B{\mathop{\rm sinc}\nolimits} \left( {\pi B{v_m}} \right).
\end{equation}
The shape of the passsband can be given by
\begin{equation}\label{eq:passband_shape}
{\left| {{R_0}(2\pi \phi {f_m})} \right|^2} = N_0^2{B^2}{{\mathop{\rm sinc}\nolimits} ^2}\left( {2{\pi ^2}B\phi {f_m}} \right).
\end{equation}
According to (\ref{eq:H}) and (\ref{eq:P_ssb}), the response can be expressed as
\begin{equation}\label{eq:P_rev}
\begin{array}{l}
P_{ssb}^{sig}({v_m}) = 2{(\gamma /2)^2} \cdot \\
{\left| {2{R_0}({v_m}) + {R_0}({v_m} - d){e^{ - j2\pi {f_0}d}} + {R_0}({v_m} + d){e^{j2\pi {f_0}d}}} \right|^2}.
\end{array}
\end{equation}
The MPF has a low-pass response and a bandpass response. The center frequency of the passband is determined by $v_m-d=0$, so the center frequency can be expressed as
\begin{equation} \label{eq:center_frequency}
{f_c} = \frac{d}{{2\pi \phi }}.
\end{equation}
According to (\ref{eq:center_frequency}), the center frequency can be tuned by adjusting the time delay of the interferometer. We suppose the bandwidth of the passband is $\Delta f$ and the bandwidth is far less than the center frequency, i.e. $\Delta f \ll f_c$, so $\Delta v = 2\pi \phi \Delta f \ll 2\pi \phi {f_c} = {v_c} = d$. If the RF signal falls into the passband, i.e. $\left| {{v_m} \pm d} \right| \leq \Delta v/2$, then ${v_m} \in \left[ { \pm d - \Delta v/2, \pm d + \Delta v/2} \right]$.
If ${v_m} \in \left[ {d - \Delta v/2,d + \Delta v/2} \right]$, then ${R_0}({v_m}) \approx {R_0}(d) \approx 0$, ${R_0}({v_m} + d) \approx {R_0}(2d) \approx 0$, and ${R_0}({v_m} - d) \approx {R_0}(0)$; if ${v_m} \in \left[ { - d - \Delta v/2, - d + \Delta v/2} \right]$, then ${R_0}({v_m}) \approx {R_0}( - d) \approx 0$, ${R_0}({v_m} - d) \approx {R_0}( - 2d) \approx 0$, and ${R_0}({v_m} + d) \approx {R_0}(0)$. In both cases, ${R_0}({v_m}){R_0}({v_m} + d) \approx 0$, ${R_0}({v_m}){R_0}({v_m} - d) \approx 0$, and ${R_0}({v_m} + d){R_0}({v_m} - d) \approx 0$. As a consequence, (\ref{eq:P_rev}) can be simplified to
\begin{equation}\label{eq:P_ssb^sig}
\begin{array}{l}
P_{ssb}^{sig}({v_m}) \approx \\
2{(\gamma /2)^2}\left[ {4{{\left| {{R_0}({v_m})} \right|}^2} + {{\left| {{R_0}({v_m} - d)} \right|}^2} + {{\left| {{R_0}({v_m} + d)} \right|}^2}} \right].
\end{array}
\end{equation}
As can be seen, the MPF for the SSB modulation has a lowpass response and a passband response without power fading effect. Furthermore, we will calculate the power of the signal falling into the passband. Specifically, the power of the RF signal whose frequency equals the center frequency of the passband will be calculated. Therefore, (\ref{eq:P_ssb^sig}) can be rewritten as
\begin{IEEEeqnarray*} {rCl}
P_{ssb}^{sig}({f_c}) &=& 2{(\gamma /2)^2})\left[ {4{{\left| {{R_0}(d)} \right|}^2} + {{\left| {{R_0}(0)} \right|}^2} + {{\left| {{R_0}(2d)} \right|}^2}} \right].\\
\IEEEyesnumber \label{eq:Pssb}
\end{IEEEeqnarray*}
Since the bandwidth of the passband is far less than the center frequency, ${{{\left| {{R_0}(d)} \right|}^2}} \approx 0$ and ${{{\left| {{R_0}(2d)} \right|}^2}} \approx 0$. Then (\ref{eq:Pssb}) can be approximated to
\begin{IEEEeqnarray} {rCl} \label{P_ssb^sig}
{P_{ssb}^{sig}}({f_c}) &\approx& 2{(\gamma/2) ^2}{\left| {{R_0}(0)} \right|^2}=2{(\gamma/2) ^2}N_0^2{B^2}.
\end{IEEEeqnarray}
On the other hand, the noise power at $\pm f_c$ within 1 Hz bandwidth is calculated by evaluating (\ref{eq:no}) at $\pm f_c$,
\begin{IEEEeqnarray*} {rCl}
\IEEEeqnarraymulticol{3}{l}{ {P_{ssb}^{no}}({f_c})=2S^{no}(f_c) }\\
&=& 2{\left| {R_{2m}^{(0)}({v_c})} \right|^2}{S_H}({f_c}) + 2{\left| {R_{2m}^{(1)}({v_c})} \right|^2}{S_H}(2{f_c}) \\
&&+ 2{\left| {R_{2m}^{( - 1)}({v_c})} \right|^2}{S_H}(0)\\
&=&2\left[ {1 + {{(\gamma /2)}^4} + 2{{(\gamma /2)}^2}\cos (2\pi {f_c}{v_c})} \right]{S_H}({f_c})\\
 &&+ 2{(\gamma /2)^2}{S_H}(2{f_c}) + 2{(\gamma /2)^2}{S_H}(0).
 \IEEEyesnumber \label{eq:Pssb_no}
\end{IEEEeqnarray*}
From Fig. {\ref{psd}} and (\ref{eq:Pssb_no}), we can see the noise power comes from the baseband noise and the up-converted noise. In (\ref{eq:Pssb_no}), $S_H(f)$ is given by
\begin{equation} \label{eq:S_H}
{S_H}(f) = \mathscr{F} \left[ {{{\left| {H(u)} \right|}^2}} \right] = \left[ {4 + 2\cos (2\pi fd)} \right]{S_0}(f),
\end{equation}
where $S_0(f)$ is expressed as
\begin{IEEEeqnarray*}{rCl}
{S_0}(f) &=& \mathscr{F}\left[ {{{\left| {{R_0}(u)} \right|}^2}} \right] = \mathscr{F}\left[ {{R_0}(u)R_0^*(u)} \right]\\
&=& G(f) \otimes G( - f) = \left\{
\begin{IEEEeqnarraybox} [] [c] {l?s}
\IEEEstrut
N_0^2\left( {B - \left| f \right|} \right) & $\left| f \right| \le B$, \\
0 & otherwise.
\IEEEstrut
\end{IEEEeqnarraybox}
\right.\IEEEyesnumber \label{eq:S0}
\end{IEEEeqnarray*}
In most of cases, the bandwidth of the OF is far larger than the frequency of the RF signal, i.e. $B \gg f_c$, and then $S_0(0)\approx S_0(f_c) \approx S_0(2f_c)$.
\par In optical links based on the IBOS, when the optical power is large, the relative intensity noise (RIN) of the IBOS is the dominant noise source \cite{Huang_JQE}. Therefore, we only need to consider the RIN for the SNR analysis. Based on the above consideration, the SNR of SSB modulation can be obtained,
\begin{IEEEeqnarray*}{l}
SNR_{out} = {P_{ssb}^{sig}(f_c)}/{P_{ssb}^{no}(f_c)} \\
\approx \frac{B}{{8{{\left[ {\cos (2\pi {f_c}{v_c}) + 1/2} \right]}^2} + 8/{\gamma ^2}\left[ {\cos (2\pi {f_c}{v_c}) + 2} \right] + 6}}.\\
\IEEEyesnumber \label{eq:snr_ssb}
\end{IEEEeqnarray*}
As can be seen from (\ref{eq:snr_ssb}), the output SNR is a function of the center frequency of the passband $f_c$, the modulation index $\gamma$, the optical bandwidth $B$, and the chromatic dispersion $\phi$. We can observe that the SNR is improved with the increase of the modulation index and proportional to the optical bandwidth of the OF. Moreover, the SNR varies periodically with the center frequency of the passband and the chromatic dispersion.
In the experiment, $\gamma$ can be obtained by measuring the carrier-to-sideband ratio (CSR). Since $10\lg {(\gamma/2) ^{-2}} = CSR$, $\gamma$ can be obtained by $\gamma  = 2\times {{{10}^{-CSR/20}}}$.

\subsection{The MPF in which only one arm is modulated}
The schematic diagram of the MPF in which only one arm is modulated is shown in Fig. \ref{setup2} as the second typical configuration. To overcome the power fading of the passband induced by the chromatic dispersion of the dispersive medium, the modulation can be phase modulation or optical carrier suppression (OCS) modulation \cite{Xue}. In Section III-B, we only give the SNR of the phase modulation. For the phase modulation, $m(t)$ can be expressed as $m(t) = {e^{j\gamma \sin (2\pi {f_m}t)}} \approx {J_0}(\gamma ) + {J_1}(\gamma ){e^{j2\pi {f_m}t}} - {J_1}(\gamma ){e^{ - j2\pi {f_m}t}}$ \cite{Xue}. Then
(\ref{eq:general_autocor}) can be simplified as,
\begin{IEEEeqnarray*} {rCl}
\IEEEeqnarraymulticol{3}{l}{
{R_{4{E_D}}}(v,v + u,u)} \\
&=& \left[ {{{\left| {{R_0}(v)} \right|}^2} \!+\! {{\left| {{R_0}(u)} \right|}^2}} \right] \\
&&+ \left[ {{{\left| {{R_0}(v)} \right|}^2} \!+\! {{\left| {{R_0}(u + d)} \right|}^2}} \right]\!\left[ {J_0^2 + 2J_1^2\cos({2\pi{f_m}}v)} \right]\\
&&+ \left[ {{{\left| {{R_0}(v + d)} \right|}^2} \!+\! {{\left| {{R_0}(u)} \right|}^2}} \right]\!\left[ {J_0^2 + 2J_1^2\cos({2\pi{f_m}}u)} \right]\\
&&+ \left[ {{{\left| {{R_0}(v - d)} \right|}^2} \!+\! {{\left| {{R_0}(u)} \right|}^2}} \right]\!\left[ {J_0^2 + 2J_1^2\cos({2\pi{f_m}}u)} \right]\\
&&+ \left[ {{{\left| {{R_0}(v)} \right|}^2} \!+\! {{\left| {{R_0}(u - d)} \right|}^2}} \right]\!\left[ {J_0^2 + 2J_1^2\cos({2\pi{f_m}}v)} \right]\\
&&+ \left[ {{{\left| {{R_0}(v)} \right|}^2} \!+\! {{\left| {{R_0}(u)} \right|}^2}} \right]\! \cdot \bigg\{ J_0^4 + 4J_0^2J_1^2\cdot\\
&& \big[\! \cos (2\pi {f_m}v) \!+\! \cos (2\pi {f_m}u) \!-\! \cos (2\pi {f_m}v)\cos (2\pi {f_m}u) \big]\! \bigg\}, \\
\IEEEyesnumber
\end{IEEEeqnarray*}
where $J_0$ and $J_1$ are short forms of $J_0(\gamma)$ and $J_1(\gamma)$. The intensity PSD is given by the Fourier transform of ${R_{4E_D}} (v,v+u,u)$, i.e.,
\begin{IEEEeqnarray*} {l}
\IEEEeqnarraymulticol{1}{l}{
S(f) = \mathscr{F} \left[ {{R_{4{E_D}}}}(v,v+u,u) \right]=}\\
\bigg\{ {\left| {R_0(v)} \right|^2}\left[ {4(J_0^{2}+1)J_1^2\cos (2\pi {f_m}v) + {{(1 + J_0^2)}^2} } \right]\\
 \quad + \left[ {{{\left| {R_0(v + d)} \right|}^2} + {{\left| {R_0(v - d)} \right|}^2}} \right]J_0^2 \bigg\} \cdot \delta (f)\\
 + \bigg\{{2J_0^2J_1^2\left| {R_0(v)} \right|^2}\left[ {1 - \cos (2\pi {f_m}v)} \right]\\
 \quad \quad + J_1^2\left[ {{{\left| {R_0(v + d)} \right|}^2} + {{\left| {R_0(v - d)} \right|}^2}}
 \right] \bigg\}\\
\cdot \bigg[ {\delta (f - {f_m}) + \delta (f + {f_m})} \bigg]\\
 +\bigg\{{{{(1 + J_0^2)}^2} + 4J_0^2J_1^2\cos (2\pi {f_m}v)}\\
 \quad + 2\cos (2\pi fd)\left[ {J_0^2 + 2J_1^2\cos (2\pi {f_m}v)} \right]
 \bigg\}\cdot {S_0}(f)\\
+2J_1^2\bigg\{ {J_0^2\left[ {1 \!-\! \cos (2\pi {f_m}v)} \right] \!+\! 1} \bigg\}\bigg[ {{S_0}(f - {f_m}) \!+\! S(f + {f_m})} \bigg]. \\
\IEEEyesnumber \label{eq:S_pm}
\end{IEEEeqnarray*}
From (\ref{eq:S_pm}), we can obtain the power of the RF signal by integrating $S(f)$ at $\pm f_m$,
\begin{IEEEeqnarray*} {rCl}
{P^{sig}}({f_m}) &\approx& 8J_0^2J_1^2{\sin ^2}(\pi {f_m}{v_m}){\left| {R_0({v_m})} \right|^2}\\
&& + 2J_1^2\left[ {{{\left| {R_0({v_m} + d)} \right|}^2} + {{\left| {R_0({v_m} - d)} \right|}^2}} \right]. \IEEEyesnumber \label{eq:P_pm}
\end{IEEEeqnarray*}
As can be seen from (\ref{eq:P_pm}), the frequency response has a lowpass response eliminated by the term ${\sin ^2}(\pi {f_m}{v_m})$ and a bandpass response. The term ${\sin ^2}(\pi {f_m}{v_m})$ comes from the phase modulation induced power fading effect \cite{Zeng}.
The power of the RF signal at the center frequency of the passband can be expressed as
\begin{equation}
{P^{sig}}({f_c}) \approx 2J_1^2{\left| {R_0(0)} \right|^2} = 2J_1^2N_0^2{B^2}.
\end{equation}
We can also get the noise power within 1 Hz bandwidth by evaluating (\ref{eq:S_pm}) at $\pm f_c$,
\begin{IEEEeqnarray*} {rCl}
{P^{no}}(f_c) &\approx& 2\bigg[ 4J_1^2{{\cos }^2}(2\pi {f_c}{v_c}) + 2J_0^2\cos(2\pi {f_c}{v_c}) \\
&&+ (1 + J_0^2)(1 + J_0^2 + 4J_1^2) \bigg]. \IEEEyesnumber
\end{IEEEeqnarray*}
As a result, the SNR can be given by
\begin{IEEEeqnarray*} {l} \label{eq:snr_pm}
SNR_{output} = {P_{sig}}({f_c})/{P_{no}}({f_c})\\
\approx \frac{B}{{2{{\left[ {\cos (2\pi {f_c}{v_c}) - 1/2} \right]}^2} + 4/{\gamma ^2}\left[ {\cos (2\pi {f_c}{v_c}) + 2} \right] + 7.5}}.\\
\IEEEyesnumber
\end{IEEEeqnarray*}
 As can be seen, the SNR is also a function of the center frequency of the passband $f_c$, the modulation index $\gamma$, the optical bandwidth $B$,  and the chromatic dispersion $\phi$. More specifically, the SNR is improved with the increase of the modulation index and proportional to the optical bandwidth of the OF. The SNR also varies periodically with the center frequency of the passband and the chromatic dispersion.

\section{Experiment}
\subsection{Experimental verification of the first typical configuration of the MPF}
An experiment based on the setup shown in Fig. 1 is performed to verify the theoretical analysis. The time delay of the ODL (General Photonics VariDelay\textsuperscript{TM}) can be tuned from 0 ps to 600 ps. Firstly, the MZM is employed in the MPF, leading to the DSB modulation. The output of the MZM is connected to a DCF module whose chromatic dispersion is -989 ps/nm. The light after the DCF module is sent to an EDFA whose gain can be tuned. After the EDFA, a wavelength selective switch (WSS) is configured to act as an OF. Finally, the PD (u\textsuperscript{2}t XPDU2120R) which has a responsivity of 0.65 A/W and a bandwidth of 50 GHz is used for optical-to-electrical conversion. A vector network analyzer (VNA) (Anritsu MS4647A) is used to acquire the frequency response of the MPF.
\par
\begin{figure}[htbp]
\centering
\includegraphics [width=2.0in] {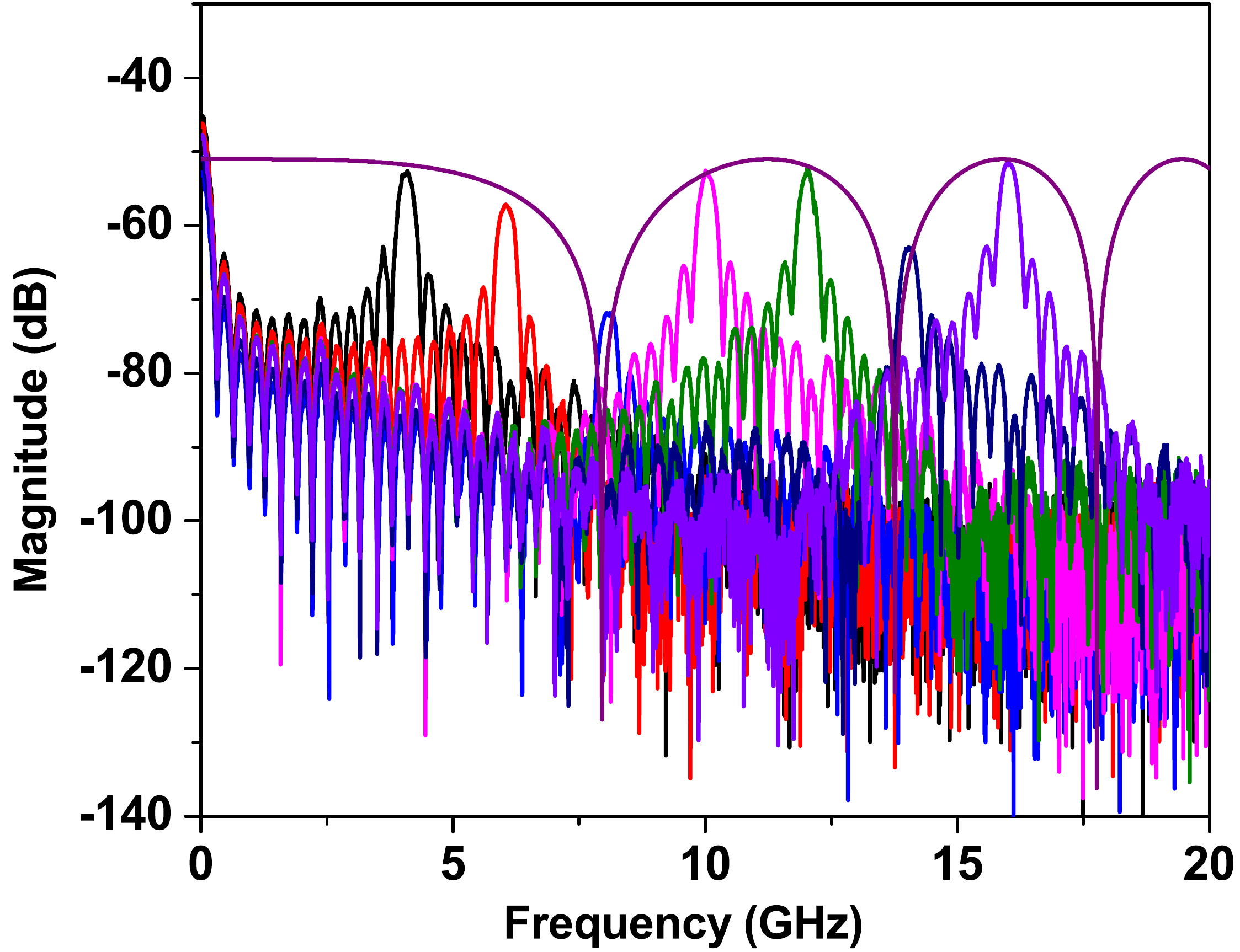}
\caption{Frequency response of the IBOS-based single-bandpass MPF using an MZM.}
\label{freResDsb}
\end{figure}

 In the experiment, the center frequency of the passband is tuned from 4 GHz to 16 GHz by adjusting the time delay of the ODL, and the experimental results are plotted in Fig. \ref{freResDsb}. As can be seen, the magnitude of passbands is shaped by the chromatic dispersion induced power fading effect, which is predicted by (\ref{eq:dsbdsb}). In particular, when the center frequency of passband is 8 GHz, the amplitude is seriously attenuated by more than 20 dB.
\begin{figure}[htbp]
\centering
\includegraphics [width=3.3in] {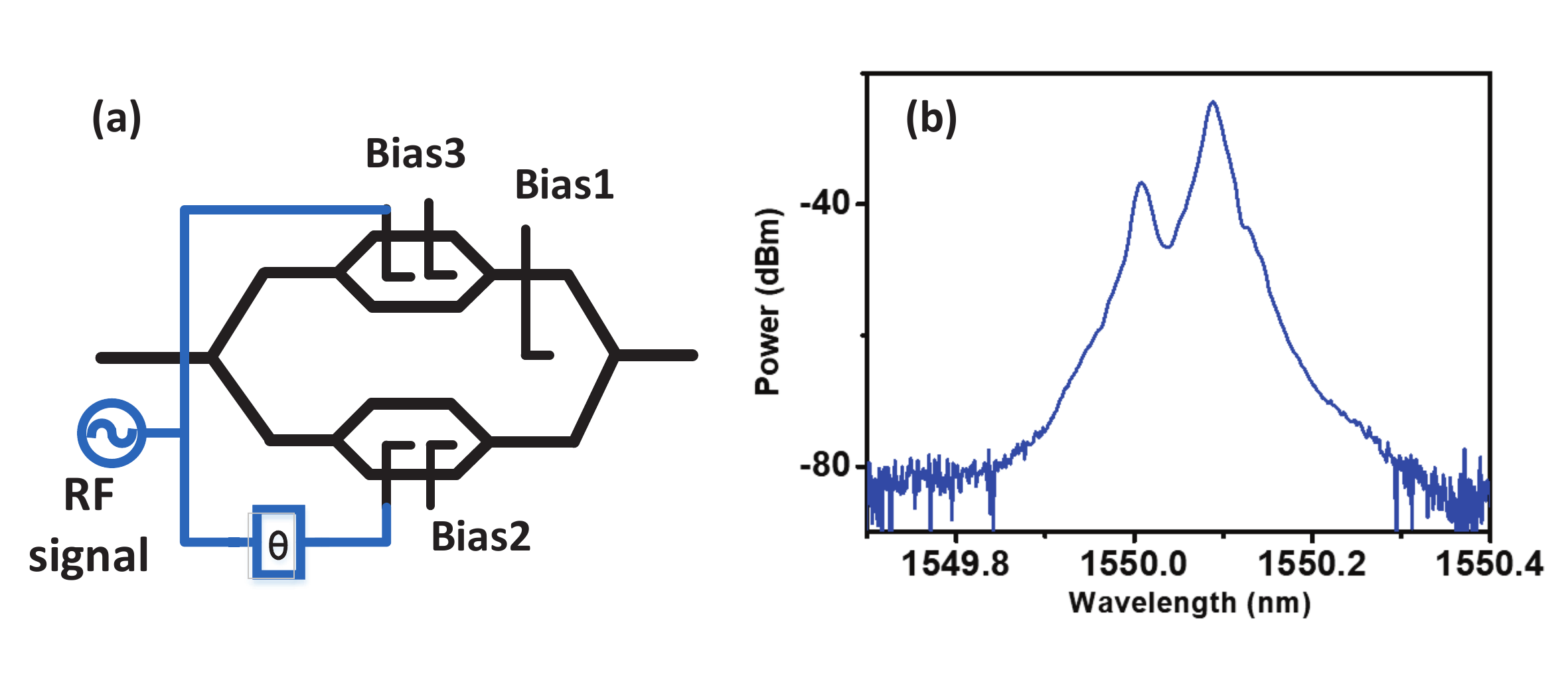}
\caption{(a) SSB modulator based on a DPMZM and (b) optical spectrum of the generated SSB signal.}
\label{ssb}
\end{figure}
\par
\begin{figure}[htbp]
\centering
\includegraphics [width=2.5in] {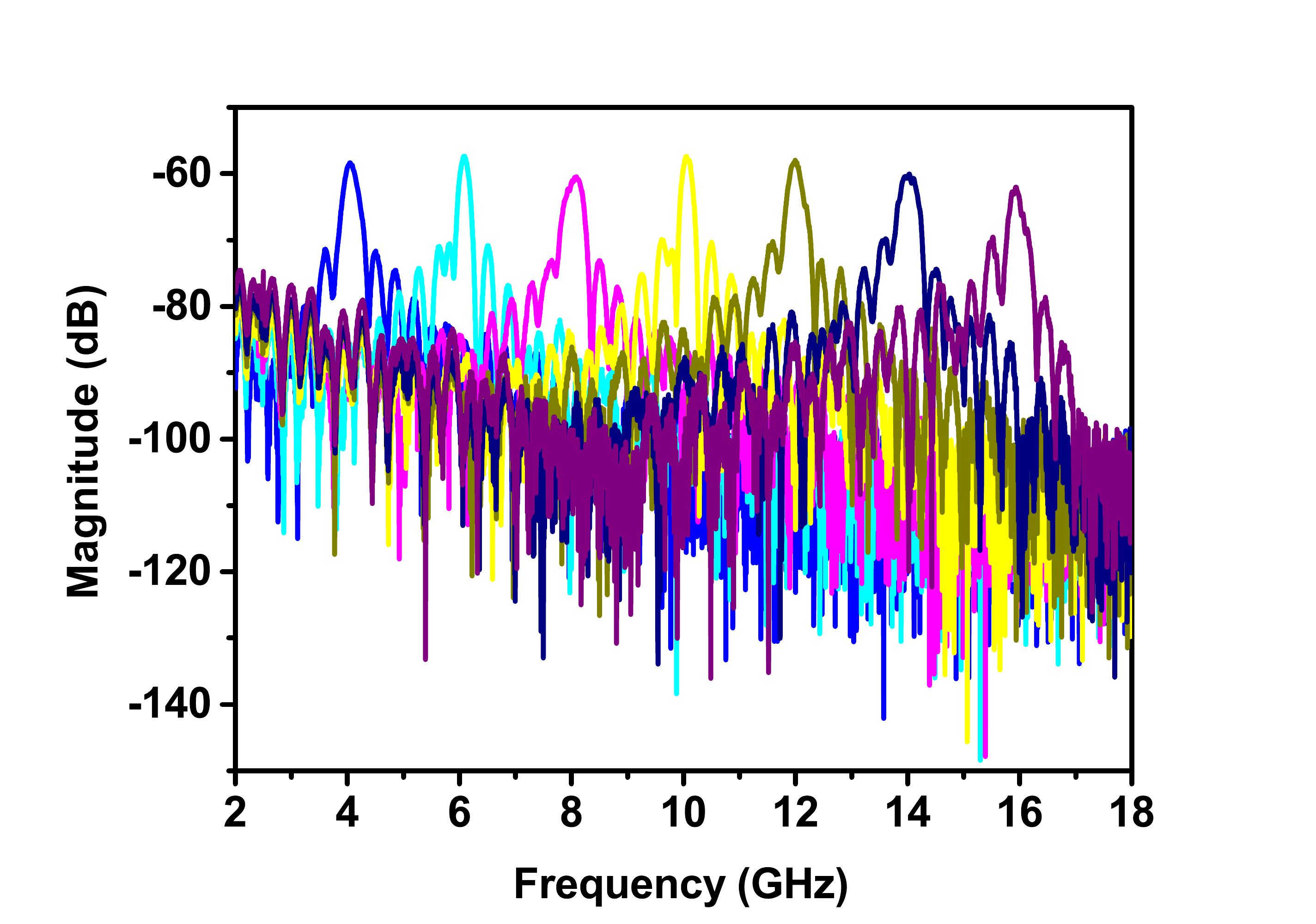}
\caption{Frequency response of the IBOS-based single-bandpass MPF using the SSB modulator.}
\label{ssbResponse}
\end{figure}

 To avoid the power fading effect, an SSB modulator should be employed. The SSB modulator can be generally achieved by a dual-drive Mach-Zehnder modulator (DDMZM) \cite{Xu, Smith} or a dual-parallel Mach-Zehnder modulator (DPMZM) \cite{Hraimel}. In our experiment, we employ the DPMZM-based SSB modulator whose schematic diagram is shown in Fig. \ref{ssb}(a). The RF signal is evenly split into two arms, and the phase of one of them is shifted by {90\degree}. Then the two RF signals are applied to the sub-MZMs of DPMZM, respectively. If the three biases of DPMZM are adjusted to the quadrature point, then an SSB signal can be generated. The optical spectrum of the SSB signal is shown in Fig. \ref{ssb}(b). The frequency response based on the SSB modulator is shown in Fig. \ref{ssbResponse}. As can be seen, the amplitude of passbands is flat. It means the dispersion induced power fading is eliminated, which is predicted by (\ref{eq:P_ssb^sig}).
\par
\begin{figure}[htbp]
\centering
\includegraphics [width=2.7in] {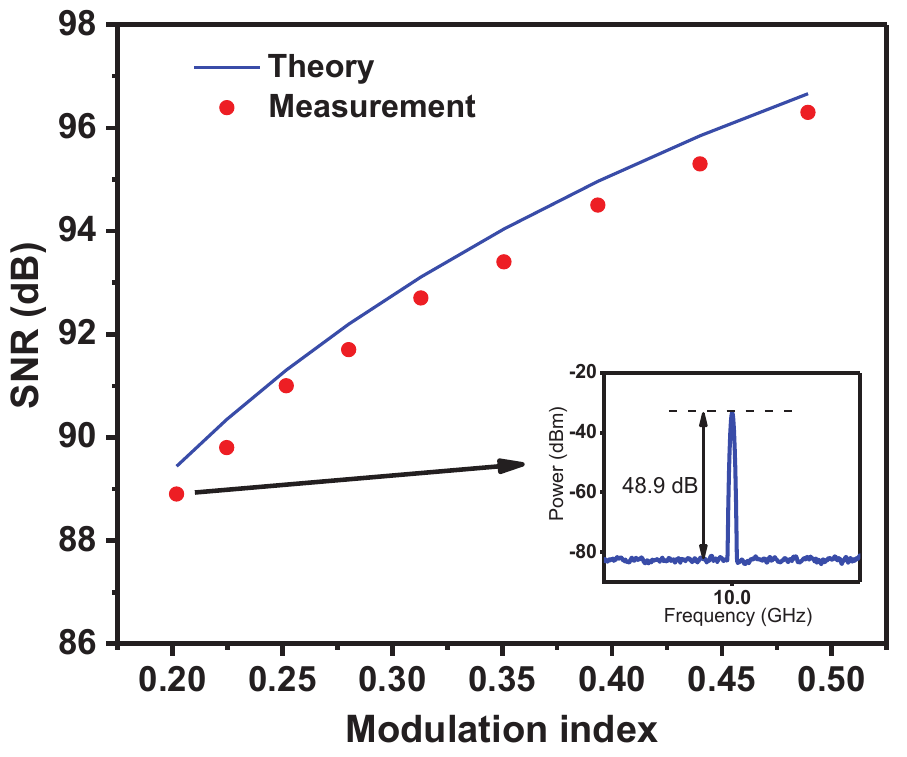}
\caption{SNR of the MPF using the SSB modulator versus modulation index. Inset: the displayed SNR is 48.9 dB because the RBW of the ESA is 10 kHz. }
\label{am_snr_mi}
\end{figure}

\begin{figure}[htbp]
\centering
\includegraphics [width=2.3in] {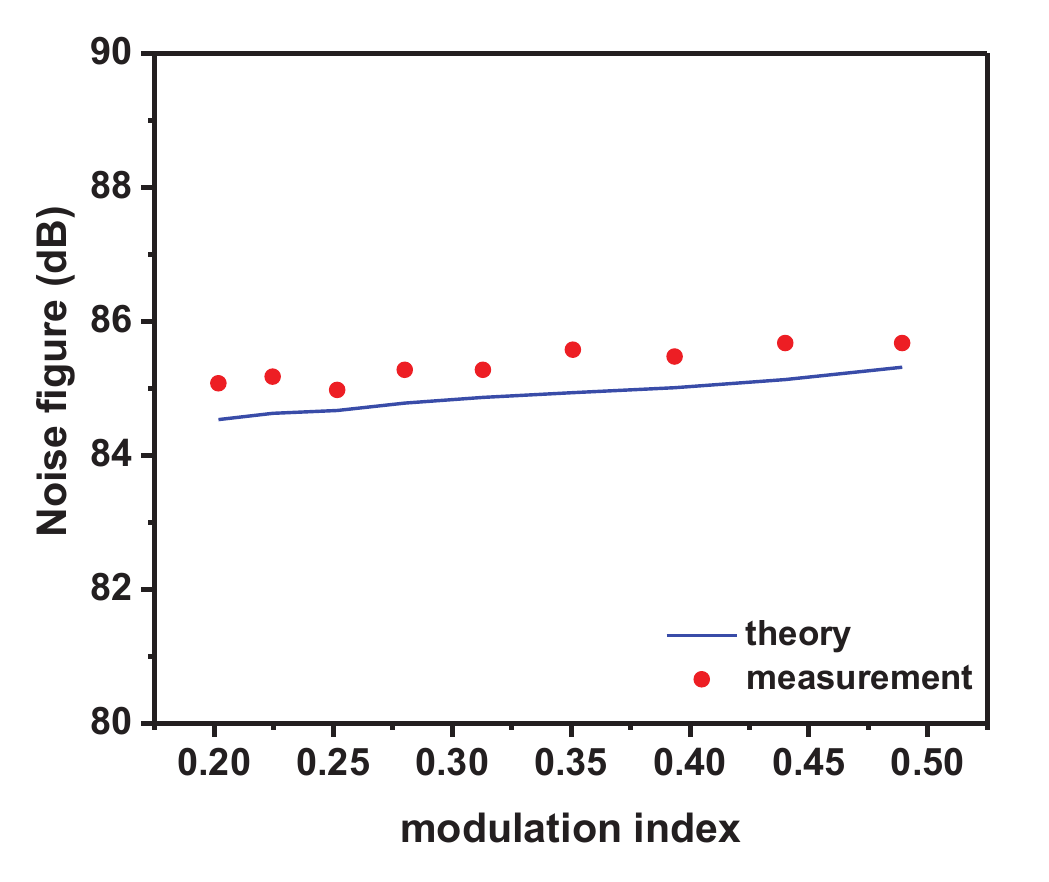}
\caption{Noise figure of the MPF using the SSB modulator versus modulation index.}
\label{nf_ssb}
\end{figure}

 Using this SSB modulator, we measure the SNR of the MPF by an electrical spectrum analyzer (ESA). In our measurement, an amplifier with a gain of 30 dB is connected to the PD to amplify the filtered signal. The resolution bandwidth (RBW) of the ESA is set to 10 kHz, so the measured value is added by 40 dB to compare with the theoretical calculation in which the noise bandwidth is set to 1 Hz. Firstly, the bandwidth of the OF keeps 3.2 nm. Subsequently, the ODL is tuned to make the center frequency of the passband to be 10 GHz. Then the frequency of the RF signal is set to 10 GHz, while the power of the RF signal is tuned to change the modulation index. The modulation index is obtained as follows. We replace the IBOS by a laser and measure the CSR at the output of the SSB modulator. Then the modulation index $\gamma$ can be calculated from measured CSR. By changing the modulation index, we can obtain the relationship between the SNR and the modulation index. We also calculate the theoretical SNR based on (\ref{eq:snr_ssb}). Both the theoretical and measured results are shown in Fig. \ref{am_snr_mi}. As can be seen, the SNR is improved along with the increase of the modulation index. Then the noise figure can be calculated by $NF =10\text{lg}[\left( {{P_{in}}/{k_B}{T_s}} \right)/SN{R_{out}}]$, where $P_{in}$ is the input RF power, $k_B$ is the Boltzmann¡¯s constant, and $T_s = 290\,\text{K}$ is the standard noise temperature. The calculated and measured noise figure is shown in Fig. \ref{nf_ssb}. As can be seen, the noise figure of the IBOS-based single-bandpass MPF is about 85 dB, which is relatively large due to the thermal noise property of the IBOS \cite{Goodman}. To reduce the noise figure, a high gain and low noise amplifier can be placed before the MPF. For example, if a amplifier with a typical gain of 30 dB and 5 dB noise figure is used, the noise figure of the cascaded MPF can be reduced to 50 dB.

\par
\begin{figure}[htbp]
\centering
\includegraphics [width=2.3in] {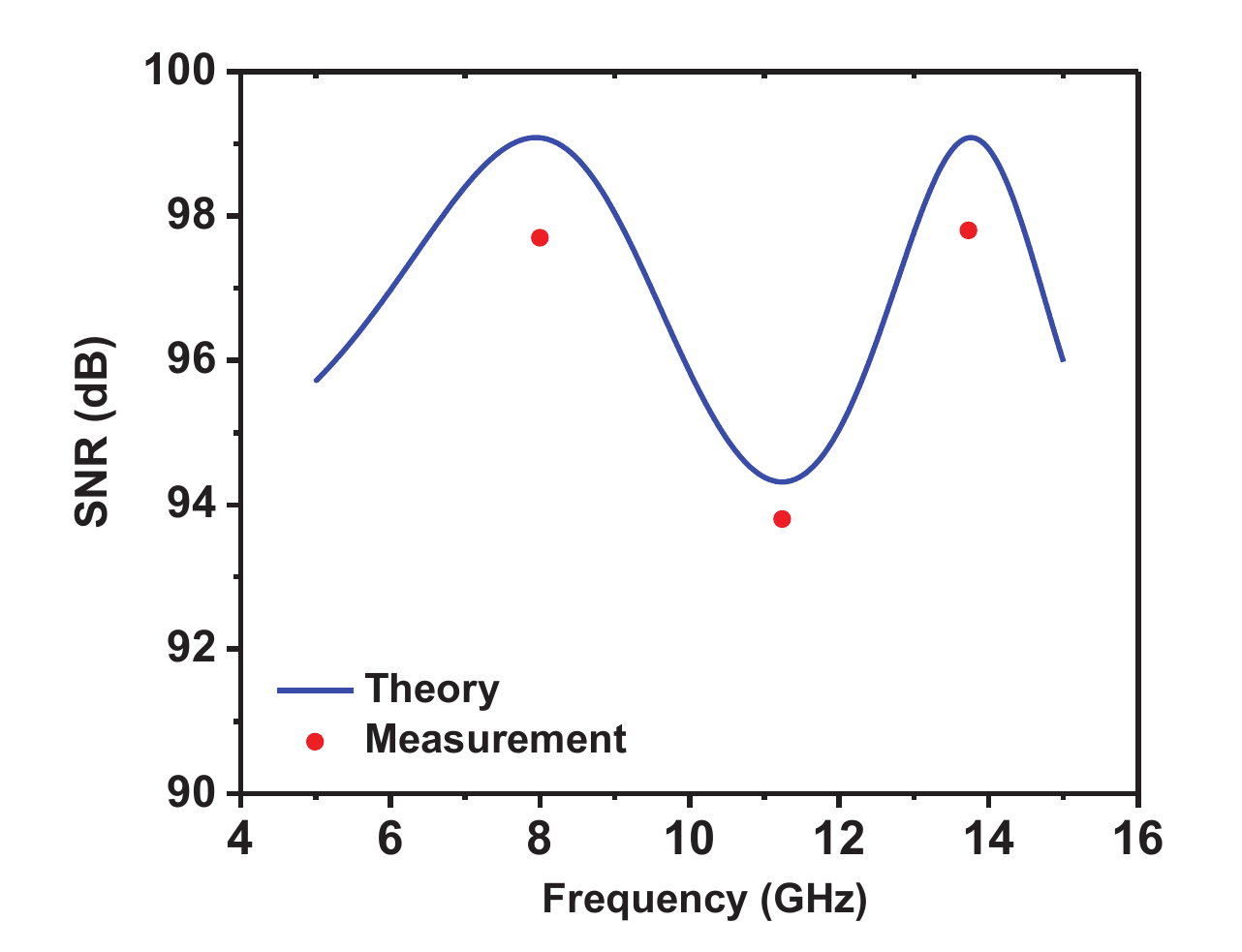}
\caption{SNR of the MPF using the SSB modulator versus frequency.}
\label{am_snr_freq}
\end{figure}

 Next, the bandwidth of OF is adjusted to 3.2 nm. The modulation index is tuned to 0.44. The ODL is adjusted to tune the center frequency of the passband, and the frequency of the RF signal is correspondingly set to the center frequency of the passband. The SNR of the filtered RF signal at each center frequency is measured. The theoretical SNR is also calculated. Both of the theoretical and measured SNRs are shown in Fig. \ref{am_snr_freq}. We can observe the SNR of different passbands is periodic rather than equal. Based on (\ref{P_ssb^sig}), the signal PSD at each frequency is flat while the noise PSD is periodic if the time delay $d$ is not zero as shown in (\ref{eq:Pssb_no}) and (\ref{eq:S_H}). Therefore, we can conclude that the periodic SNR is caused by the time delay originated from the ODL. As can be seen from Figs. \ref{am_snr_mi}-\ref{am_snr_freq}, the measured SNR basically agrees with the theoretical SNR.
\par
\begin{figure}[htbp]
\centering
\includegraphics [width=2.3in] {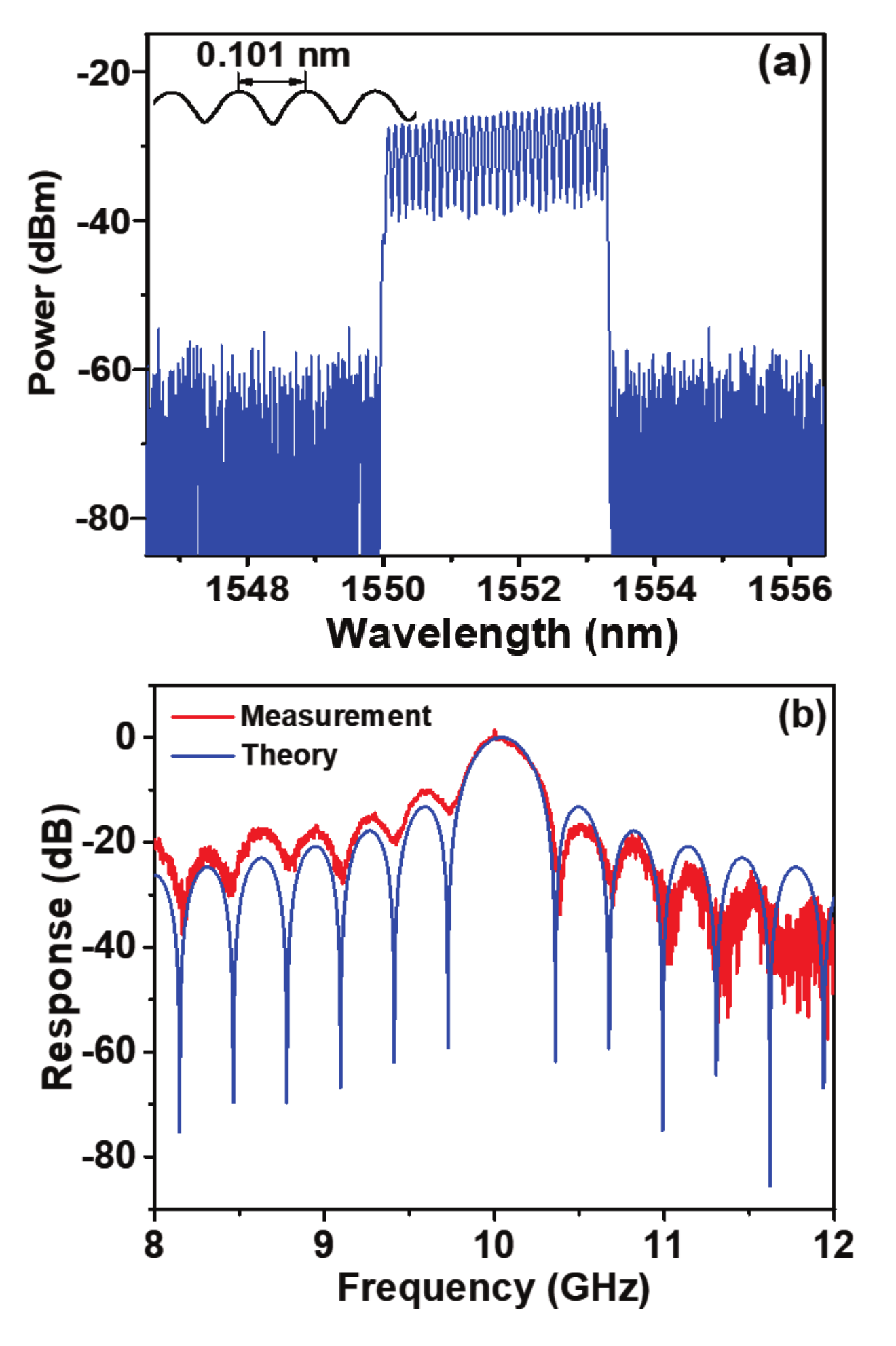}
\caption{(a) Optical spectrum and (b) frequency response of the MPF using the SSB modulator when the optical bandwidth is 3.2 nm.}
\label{ssb32}
\end{figure}

\begin{figure}[htbp]
\centering
\includegraphics [width=2.3in] {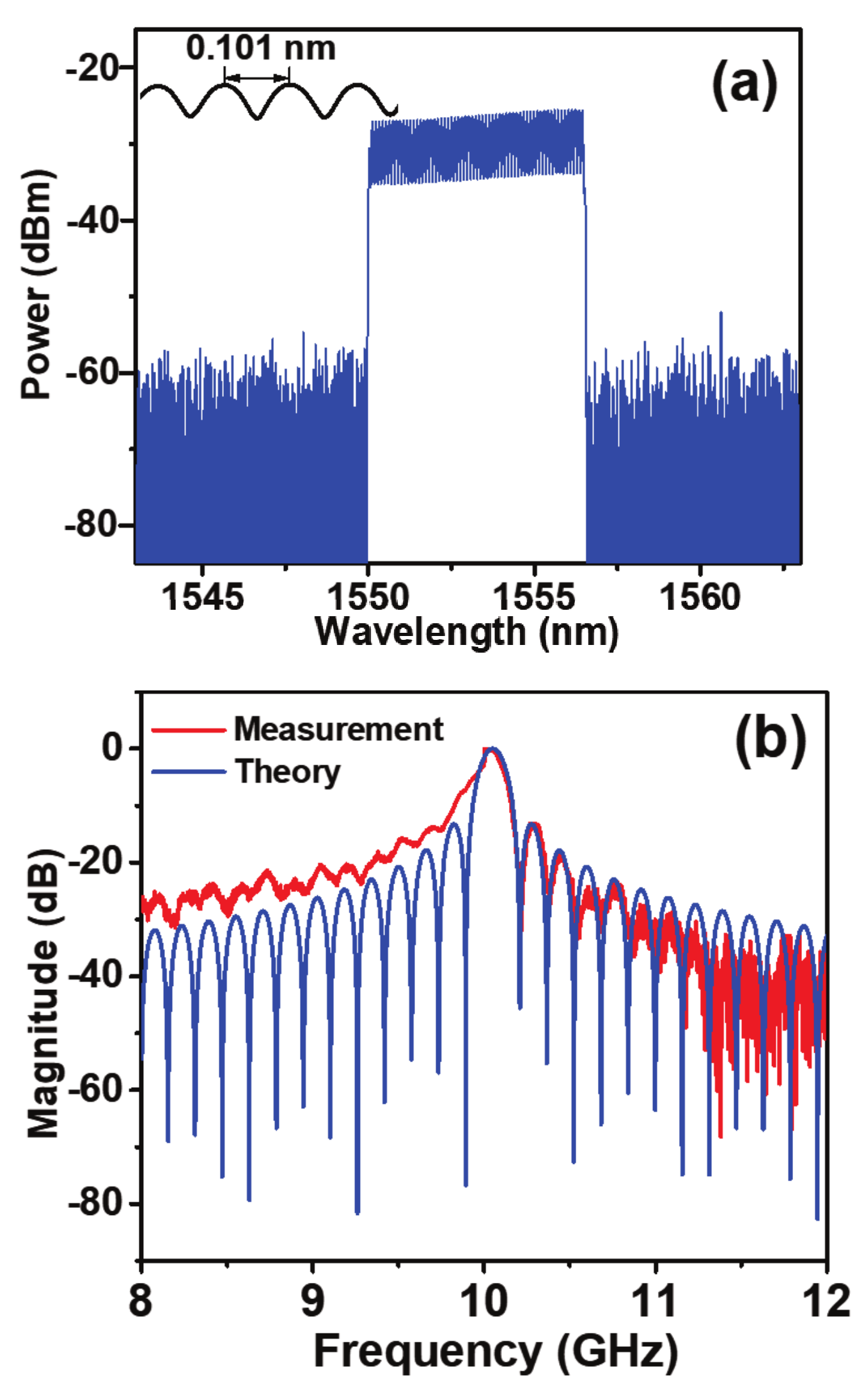}
\caption{(a) Optical spectrum and (b) frequency response of the MPF using the SSB modulator when the optical bandwidth is 6.4 nm.}
\label{ssb64}
\end{figure}

 According to (\ref{eq:passband_shape}), we can change the shape of the passband by adjusting the shape of the IBOS. The center frequency of the passband is set to 10 GHz, while the frequency of the RF signal and the modulation index are set to 10 GHz and 0.39, respectively. The bandwidth of the OF is adjusted to 3.2 nm and 6.4 nm, successively. The FSR of the optical spectrum is given by $FSR = {\lambda ^2}/(cd)$. In our experiment, the delay is tuned to 79.4 ps, so the FSR is calculated to be 0.101 nm. The optical spectra before the PD are shown in Figs. \ref{ssb32}(a) and \ref{ssb64}(a), respectively. The corresponding simulated and measured shapes of the passband are plotted in Figs. \ref{ssb32}(b) and \ref{ssb64}(b), respectively. We can see the experiment results match well with the theory. Moreover, the SNR is proportional to the bandwidth of the OF based on (\ref{eq:snr_ssb}). When the bandwidth is 3.2 nm, the measured SNR is 93.8 dB. When the bandwidth is 6.4 nm, the measured SNR is 96.6 dB. On the other hand, the calculated SNRs are 94.9 dB and 97.9 dB for 3.2 nm and 6.4 nm bandwidth, respectively. We can see when the optical bandwidth is doubled, the SNR increases by 3 dB. Therefore, the experiment results agree with the theoretical analysis, which is the SNR is proportional to the optical bandwidth.

\subsection{Experimental verification of the second typical configuration of the MPF}
\begin{figure}[htbp]
\centering
\includegraphics [width=2.5in] {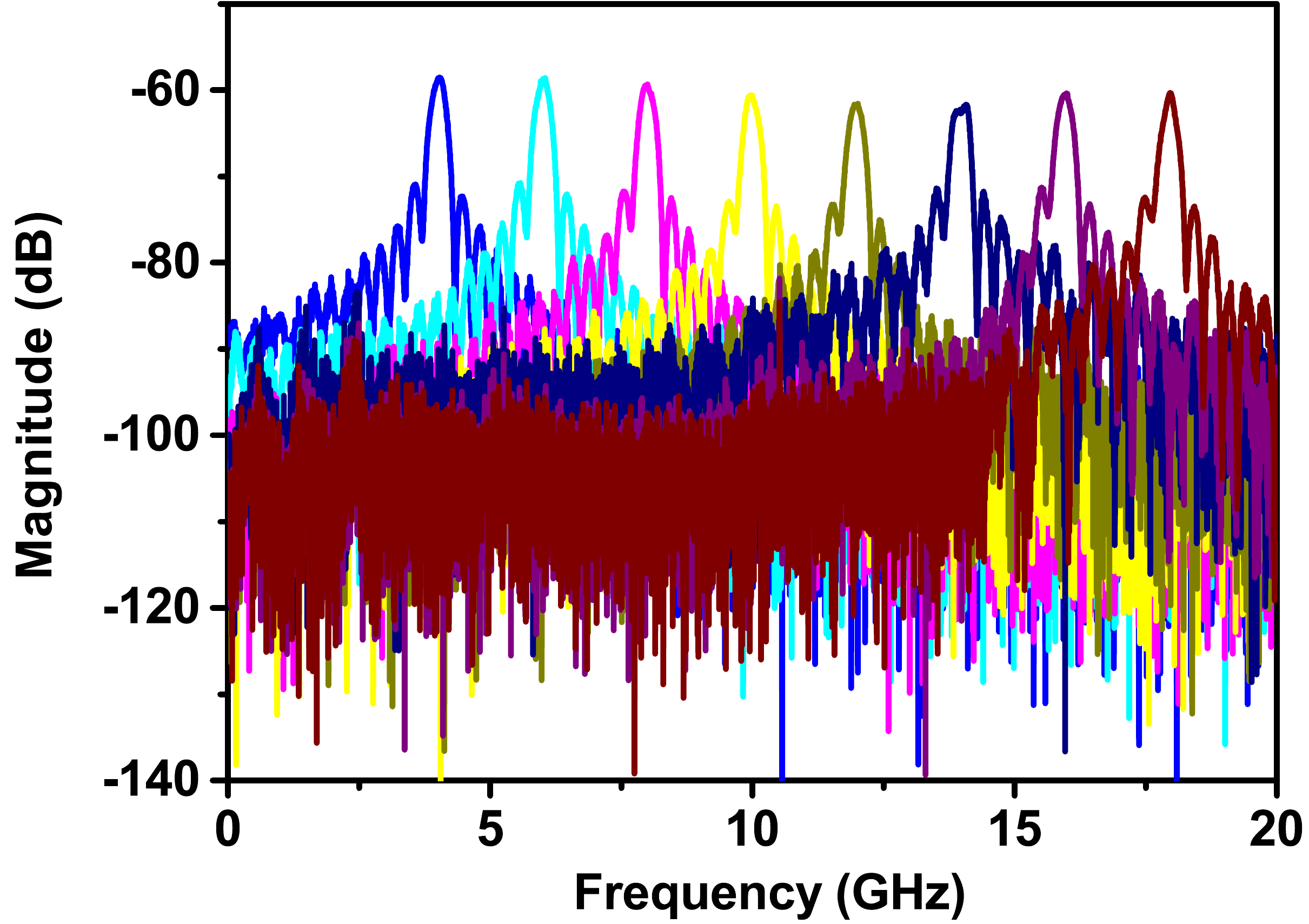}
\caption{Frequency response of the MPF using a phase modulator.}
\label{pmResponse}
\end{figure}

The second typical configuration using a phase modulator is shown in Fig. \ref{setup2}. The frequency response of the MPF is also acquired by the VNA. By adjusting the ODL, the center frequency of passband is tuned from 4 GHz to 16 GHz, and the acquired amplitude response is shown in Fig. \ref{pmResponse}. As can be seen, the amplitude of passbands is flat, which agrees with (\ref{eq:P_pm}).
\par
\begin{figure}[htbp]
\centering
\includegraphics [width=2.3in] {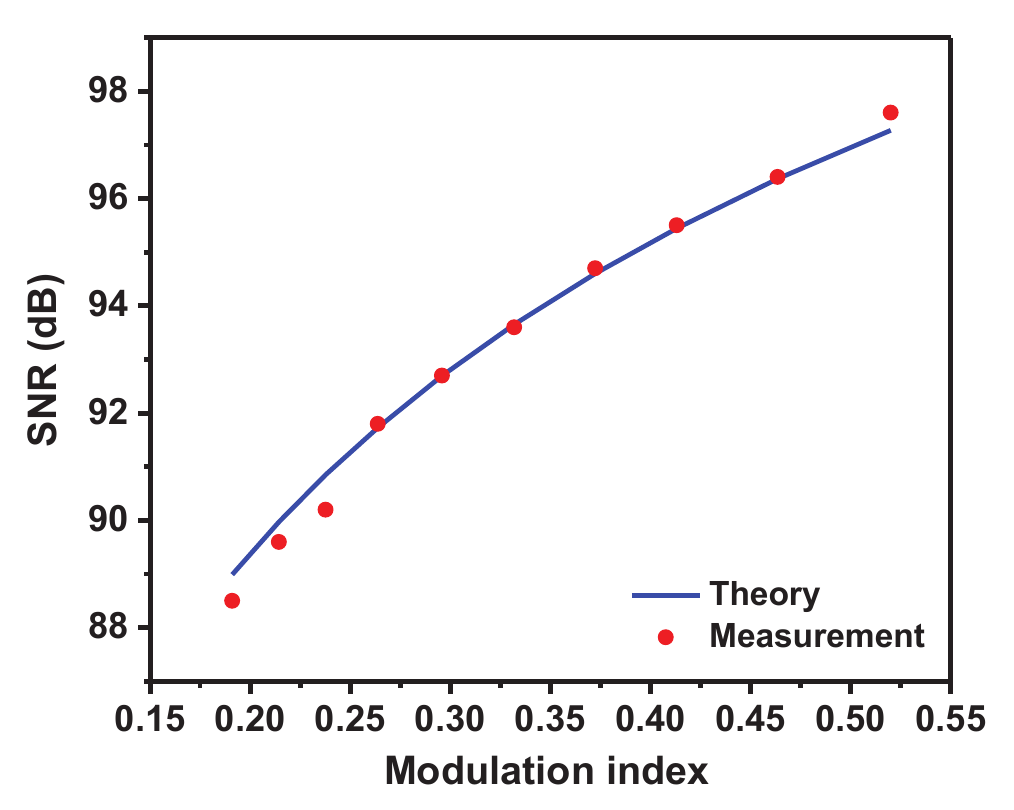}
\caption{SNR of the MPF using a phase modulator versus modulation index.}
\label{pm_snr_mi}
\end{figure}

\begin{figure}[htbp]
\centering
\includegraphics [width=2.2in] {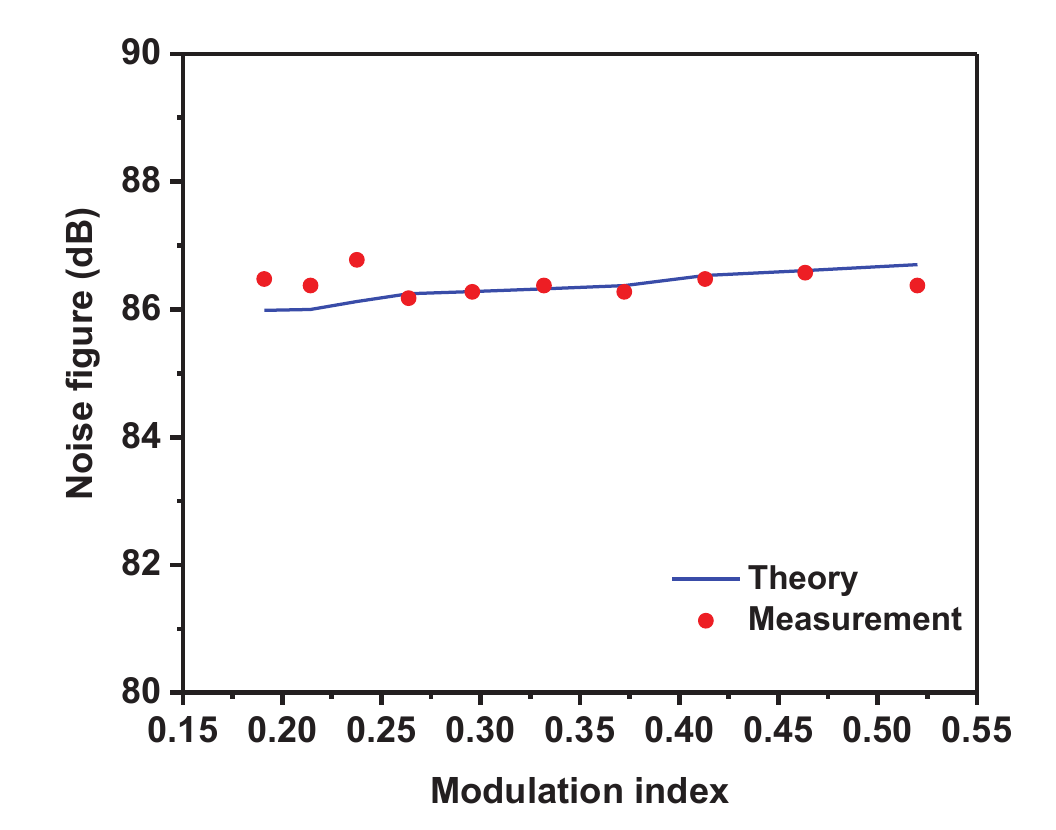}
\caption{Noise figure of the MPF using a phase modulator versus modulation index.}
\label{nf_pm}
\end{figure}

 Then the relationship between the SNR and the modulation index is measured. The bandwidth of the OF and the frequency of the RF signal are set to 3.2 nm and 10 GHz, respectively. We change the modulation index and measure the corresponding SNR. The measured SNR is shown in Fig. \ref{pm_snr_mi} where the theoretical SNR is also plotted. As can be seen, the SNR is improved along with the increase of the modulation index. Furthermore, the measured and calculated noise figure are shown in Fig. \ref{nf_pm}.

\par
\begin{figure}[htbp]
\centering
\includegraphics [width=2.0in] {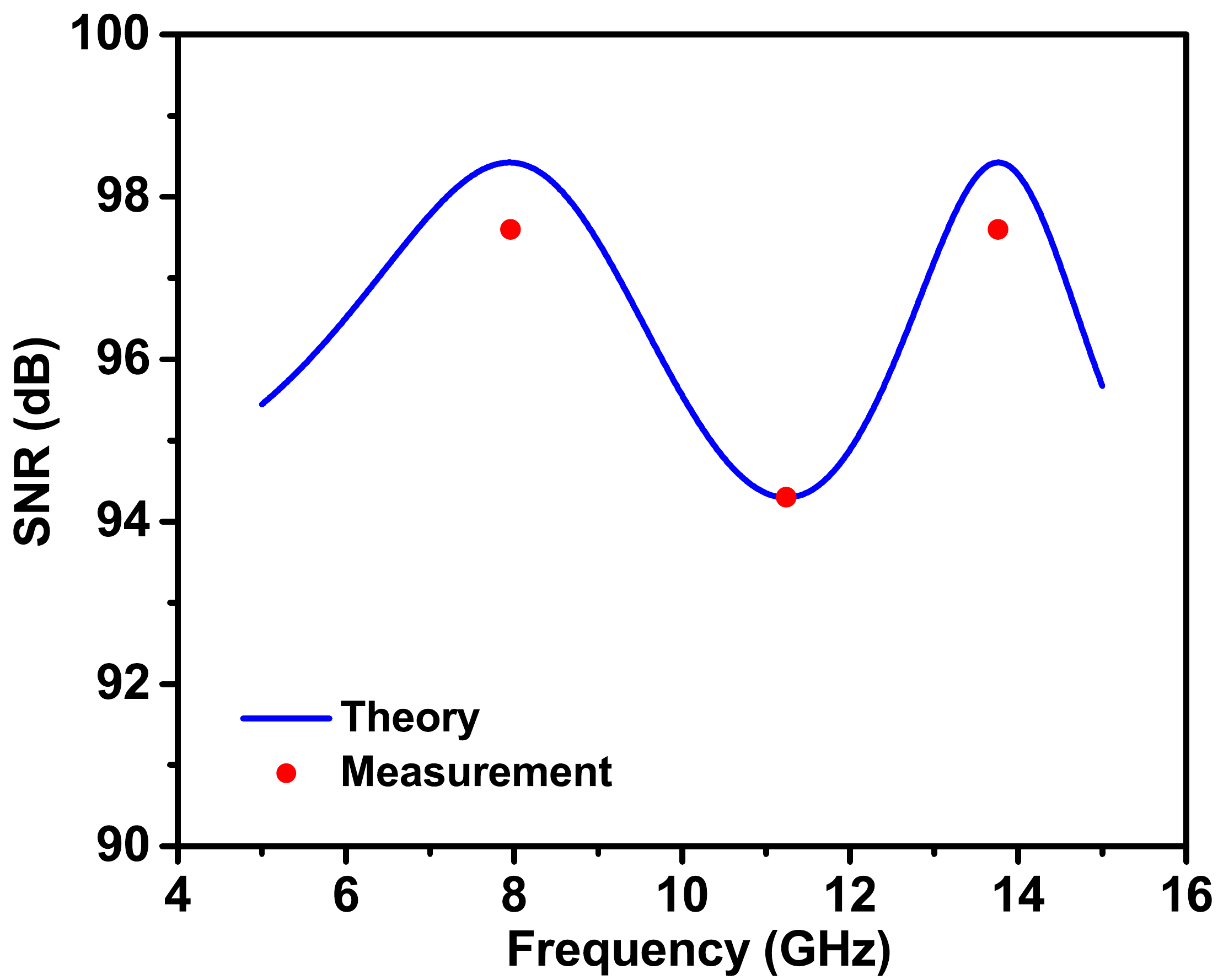}
\caption{SNR of the MPF using a phase modulator versus frequency.}
\label{pm_snr_freq}
\end{figure}

 Afterwards, the relationship between the SNR and the frequency is measured. The measurement procedure is similar to the SSB modulator-based MPF given in Section IV-A. The measured and simulated results are shown in Fig. \ref{pm_snr_freq}. We can observe that the SNR varies with the frequency periodically, again. As can be seen from Figs. \ref{pm_snr_mi}-\ref{pm_snr_freq}, the measured results agree well with the theoretical results.
\par
\begin{figure}[htbp]
\centering
\includegraphics [width=2.2in] {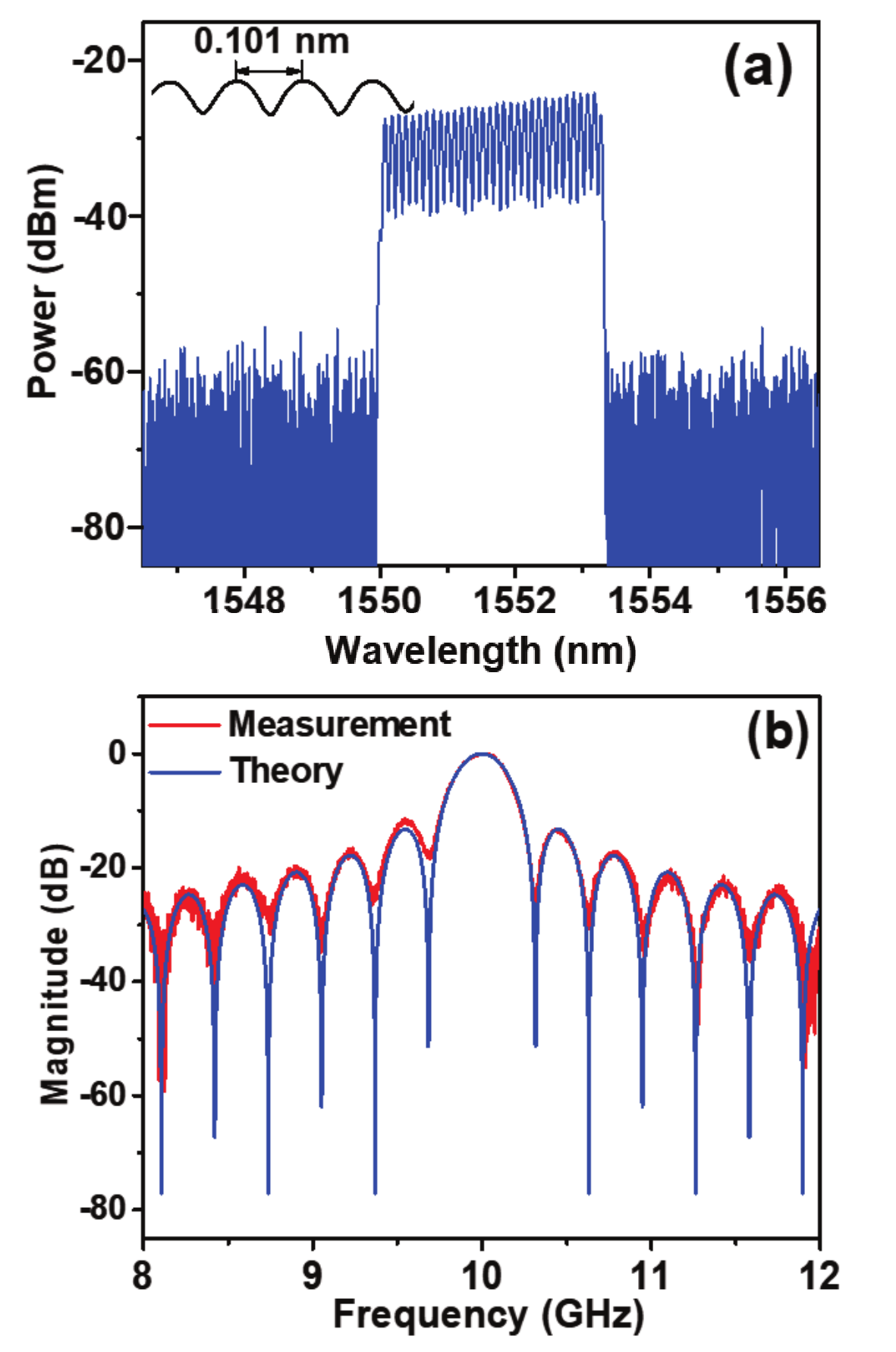}
\caption{(a) Optical spectrum and (b) frequency response of the MPF using a phase modulator when the optical bandwidth is 3.2 nm.}
\label{pm32}
\end{figure}

\begin{figure}[htbp]
\centering
\includegraphics [width=2.2in] {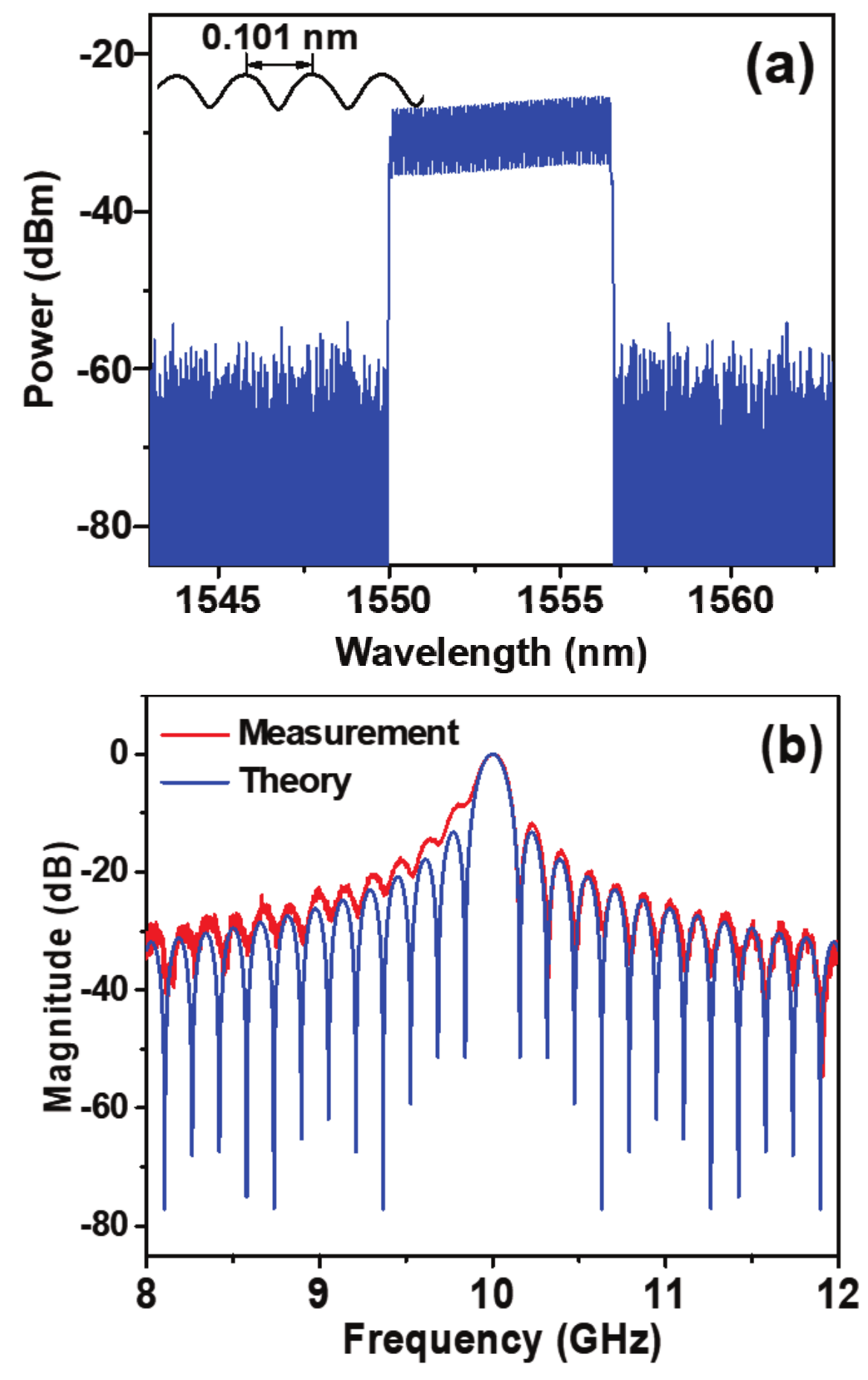}
\caption{(a) Optical spectrum and (b) frequency response of the MPF using a phase modulator when the optical bandwidth is 6.4 nm.}
\label{pm64}
\end{figure}

 Finally, we adjust the center frequency of the passband to 10 GHz by setting the time delay to 79.4 ps. Therefore, the calculated FSR is 0.101 nm. When the bandwidth of the OF is set to 3.2 nm, the measured and simulated shapes of the passband according to (\ref{eq:passband_shape}) are shown in Fig. \ref{pm32}. Furthermore, when the bandwidth of the OF is set to 6.4 nm, the measured and simulated shapes of the passband are shown in Fig. \ref{pm64}. A RF signal with a frequency of 10 GHz is applied to the modulator in which the modulation index is set to 0.41. According to (\ref{eq:snr_pm}), the SNR is proportional to the bandwidth of the OF. When the bandwidth is 3.2 nm, the measured SNR is 95.3 dB. When the bandwidth is 6.4 nm, the measured SNR is 98.1 dB. On the other hand, the simulated SNRs are 95.6 dB and 98.6 dB for 3.2 nm and 6.4 nm bandwidth, respectively. We can see when the bandwidth of the OF is doubled, the SNR increases by 3 dB. Therefore, experiment results also verify the theoretical analysis, which is the SNR is proportional to the optical bandwidth.

\section{Conclusion}
In conclusion, the output SNR of the MPF with an interferometric structure based on an IBOS was studied in this paper. A general model which represents previous configurations in the literature was proposed. Based on the general model, an analysis framework of the SNR was given. The theoretical analysis shows that the SNR is a function of the center frequency of the passband, the modulation index, the chromatic dispersion and the bandwidth of the OF. The SNR of two special cases which can be experimentally verified in our lab was investigated in detail. An experiment was performed to verify the analysis, and the results agree well with the theoretical calculation. It is worth noting that the interferometric structure in the MPF analyzed in this paper has two arms. If the interferometric structure has more arms, more passbands will be generated \cite{Huang, Ge}. Our analysis method can be easily extended to analyze more arms configuration. We believe the analysis of the SNR will help evaluate the performance of the applications based on this kind of MPFs, such as RoF systems and OEOs.

\section*{Appendix I}
The analysis presented in Section III is based on solving the field envelope function over the dispersive medium, i.e. (\ref{eq:dispersive}), in the time domain. Generally speaking, the frequency domain analysis is more convenient when dealing with chromatic dispersion of the dispersive medium. In this section, the SNR of the MPF using an SSB modulator is calculated by the frequency domain analysis, and we obtain the same results as given in Section III-A.
\par The frequency components of the IBOS ${{\tilde E}_0}(f)$ can be given by applying Fourier transform to ${E_0}(t)$, i.e.,

\begin{equation}
{\tilde E_0}(f) = \int\limits_{ - \infty }^{ + \infty } {{E_0}(t){e^{ - j2\pi ft}}dt}.
\end{equation}
The integral is interpreted as a mean square limit \cite{Papoulis}. Since ${E_0}(t)$ is a stationary process, the autocorrelation of ${{\tilde E}_0}(f)$ can be expressed as\cite{Papoulis}
\begin{IEEEeqnarray*} {rCl}
\left\langle\! {{{\tilde E}_0}(f)\tilde E_0^*(f')} \!\right\rangle  &=& \int\limits_{ - \infty }^{ + \infty } \! {\int\limits_{ - \infty }^{ + \infty }\! {\left\langle {{E_0}(t)E_0^*(t')} \right\rangle } {e^{ - j2\pi ft}}{e^{j2\pi f't'}}\!dtdt'} \\
 &=& \int\limits_{ - \infty }^{ + \infty } \! {\int\limits_{ - \infty }^{ + \infty } {R_0(\tau ){e^{ - j2\pi (f - f')t'}}{e^{ - j2\pi f\tau }}d\tau dt'} } \\
 &=& G(f')\delta (f - f')\IEEEyesnumber \label{eq:xcor_freq}.
\IEEEyesnumber
\end{IEEEeqnarray*}
 As can be seen, the frequency components of the IBOS at different frequencies are uncorrelated. The fourth-order moments of the frequency components of ${{\tilde E}_0}(f)$ can be derived from the properties of the time domain signal, i.e.,
\begin{IEEEeqnarray*} {rCl}
\IEEEeqnarraymulticol{3}{l}{
\left\langle {{{\tilde E}_0}({f_1})\tilde E_0^*({f_2}){{\tilde E}_0}({f_3})\tilde E_0^*({f_4})} \right\rangle}\\
 &=& \left\langle {\int\limits_{ - \infty }^{ + \infty } {{E_0}({t_1}){e^{ - j2\pi {f_1}{t_1}}}d{t_1}} \int\limits_{ - \infty }^{ + \infty } {E_0^*({t_2}){e^{j2\pi {f_2}{t_2}}}d{t_2}} } \right.\\
&&\left. {\int\limits_{ - \infty }^{ + \infty } {{E_0}({t_3}){e^{ - j2\pi {f_3}{t_3}}}d{t_3}} \int\limits_{ - \infty }^{ + \infty } {E_0^*({t_4}){e^{j2\pi {f_4}{t_4}}}d{t_4}} } \right\rangle \\
 &=& \int\limits_{ - \infty }^{ + \infty }\!{\int\limits_{ - \infty }^{ + \infty }\! {\int\limits_{ - \infty }^{ + \infty } {\int\limits_{ - \infty }^{ + \infty }\! {\left\langle {{E_0}({t_1})E_0^*({t_2}){E_0}({t_3})E_0^*({t_4})} \right\rangle } } } } \\
&&{e^{ - j2\pi {f_1}{t_1}}}{e^{j2\pi {f_2}{t_2}}}{e^{ - j2\pi {f_3}{t_3}}}{e^{j2\pi {f_4}{t_4}}}d{t_1}d{t_2}d{t_3}d{t_4}\\
 &=& \int\limits_{ - \infty }^{ + \infty } {\int\limits_{ - \infty }^{ + \infty } {\int\limits_{ - \infty }^{ + \infty } {\int\limits_{ - \infty }^{ + \infty } {\left\langle {{E_0}({t_1})E_0^*({t_2}))} \right\rangle } } } } \left\langle {{E_0}({t_3})E_0^*({t_4})} \right\rangle \\
&&{e^{ - j2\pi {f_1}{t_1}}}{e^{j2\pi {f_2}{t_2}}}{e^{ - j2\pi {f_3}{t_3}}}{e^{j2\pi {f_4}{t_4}}}d{t_1}d{t_2}d{t_3}d{t_4}\\
 &&+ \int\limits_{ - \infty }^{ + \infty } {\int\limits_{ - \infty }^{ + \infty } {\int\limits_{ - \infty }^{ + \infty } {\int\limits_{ - \infty }^{ + \infty } {\left\langle {{E_0}({t_1})E_0^*({t_4}))} \right\rangle } } } } \left\langle {{E_0}({t_3})E_0^*({t_2})} \right\rangle \\
 &=& \left\langle {{{\tilde E}_0}({f_1})\tilde E_0^*({f_2})} \right\rangle \left\langle {{{\tilde E}_0}({f_3})\tilde E_0^*({f_4})} \right\rangle \\
 &&+ \left\langle {{{\tilde E}_0}({f_1})\tilde E_0^*({f_4})} \right\rangle \left\langle {{{\tilde E}_0}({f_3})\tilde E_0^*({f_2})} \right\rangle.
\IEEEyesnumber \label{eq:4moment}
\end{IEEEeqnarray*}
It is worth noting that (\ref{eq:xcor_freq}) and (\ref{eq:4moment}) are given in \cite{Duan} by solving the Helmholtz propagation equation. However, in this paper, the frequency-domain properties of the IBOS is derived from the well-established time domain properties \cite{Goodman}. The optical field before the PD can be expressed as
\begin{IEEEeqnarray*} {rCl}
\IEEEeqnarraymulticol{3}{l}{
{E_{pd}}(t) = {E_D}(t) \otimes \eta (t)}\\
 = \left[ {{E_0}(t) + {E_0}(t - d){e^{ - j2\pi {f_0}d}}} \right]m(t) \otimes \eta (t),
\IEEEyesnumber
\end{IEEEeqnarray*}
where $\eta(t)$ is the impulse response of the dispersive medium. Then the frequency components $E_{pd}(f)$ can be given by
\begin{IEEEeqnarray*} {rCl}
{{\tilde E}_{pd}}(f) &=& \mathscr{F}\left[ {{E_{pd}}(t)} \right]\\
&{\rm{ = }}&\left[ {{{\tilde E}_0}(f)(1 + {e^{ - j2\pi (f + {f_0})d}})} \right.{\rm{ + }}\\
&&\left. {(\gamma /2){{\tilde E}_0}(f - {f_m})\left( {1 + {e^{ - j2\pi (f - {f_m} + {f_0})d}}} \right)} \right]T(f),\\
\IEEEyesnumber
\end{IEEEeqnarray*}
where $T(f) = \mathscr{F}\left[ {\eta (t)} \right]$ is the frequency response of the dispersive medium, which can be expressed as
\begin{equation}
T(f) = {e^{ - j\phi {{(2\pi f)}^2}/2}}.
\end{equation}
A benefit of the frequency domain analysis is that the higher-order dispersion of the dispersive medium can be easily included in the formula. Then the time averaged autocorrelation of the optical intensity can be written as
\begin{IEEEeqnarray*} {rCl}
\IEEEeqnarraymulticol{3}{l}{
{R_I}(\tau)}\\
 &=& \overline {{R_I}(t + \tau ,t)}\\
 &=& \overline {\left\langle {{E_{pd}}(t)E_{pd}^*(t){E_{pd}}(t + \tau )E_{pd}^*(t + \tau )} \right\rangle } \\
 &=& \overline {\left\langle {\int\limits_{ - \infty }^{ + \infty } {{{\tilde E}_{pd}}({f_1}){e^{j2\pi {f_1}t}}d{f_1}} \int\limits_{ - \infty }^{ + \infty } {\tilde E_{pd}^*({f_2}){e^{ - j2\pi {f_2}t}}d{f_2}} } \right.}\\
&&\overline {\left. {\int\limits_{ - \infty }^{ + \infty } {{{\tilde E}_{pd}}({f_3}){e^{j2\pi {f_3}(t + \tau )}}d{f_3}} \int\limits_{ + \infty }^{ + \infty } {\tilde E_{pd}^*({f_4}){e^{ - j2\pi {f_4}(t + \tau )}}d{f_4}} } \right\rangle }\\
&=& \overline {\int\limits_{ - \infty }^{ + \infty } {\int\limits_{ - \infty }^{ + \infty } {\int\limits_{ - \infty }^{ + \infty } {\int\limits_{ - \infty }^{ + \infty } {\left\langle {{{\tilde E}_{pd}}({f_1})\tilde E_{pd}^*({f_2}){{\tilde E}_{pd}}({f_3})\tilde E_{pd}^*({f_4})} \right\rangle } } } } } \\
&&\overline {{e^{j2\pi \left[ {({f_1} - {f_2})t + ({f_3} - {f_4})(t + \tau )} \right]}}d{f_1}d{f_2}d{f_3}d{f_4}}\\
&=& R_I^{sig}(\tau) + R_I^{no}(\tau),
\IEEEyesnumber \label{eq:R_I}
\end{IEEEeqnarray*}
where $R_I^{sig}(\tau)$ is the time averaged autocorrelation function of the signal, and $R_I^{no}(\tau)$ is the time averaged autocorrelation function of the noise. We apply the frequency-domain properties (\ref{eq:xcor_freq}) and (\ref{eq:4moment}) to (\ref{eq:R_I}), and then the autocorrelation function of the intensity signal can be given by
\begin{IEEEeqnarray*} {rcl}
\IEEEeqnarraymulticol{3}{l}{
R_I^{sig}(\tau)}\\
 &=& \overline {\int\limits_{ - \infty }^{ + \infty }\! {\int\limits_{ - \infty }^{ + \infty }\! {\int\limits_{ - \infty }^{ + \infty }\! {\int\limits_{ - \infty }^{ + \infty } \!\!\!{\big\langle {{{\tilde E}_0}({f_1})\tilde E_0^*({f_2} - {f_m})} \big\rangle \big\langle {{\tilde E_0}({f_3} - {f_m})\tilde E_0^*({f_4})} \big\rangle} }  }} } \\
&&\overline { \cdot \left[ {1 + {e^{ - j2\pi ({f_1} + {f_0})d}}} \right]\left[ {1 + {e^{j2\pi ({f_2} + {f_0} - {f_m})d}}} \right]} \\
&&\overline {\left[ {1 + {e^{ - j2\pi ({f_3} + {f_0} - {f_m})d}}} \right]\left[ {1 + {e^{j2\pi ({f_4} + {f_0})d}}} \right]} \\
&&\overline { \cdot {e^{j2\pi \left[ {({f_1} - {f_2})t + ({f_3} - {f_4})(t + \tau )} \right]}}d{f_1}d{f_2}d{f_3}d{f_4}}  \cdot {(\gamma /2)^2}\\
 &+&  \overline {\int\limits_{ - \infty }^{ + \infty }\! {\int\limits_{ - \infty }^{ + \infty }\! {\int\limits_{ - \infty }^{ + \infty }\! {\int\limits_{ - \infty }^{ + \infty }\!\!\! {\big\langle {{{\tilde E}_0}({f_1} - {f_m})\tilde E_0^*({f_2})} \big\rangle \big\langle {{{\tilde E}_0}({f_3})\tilde E_0^*({f_4} - {f_m})} \big\rangle} } } } } \\
&&\overline { \cdot \left[ {1 + {e^{ - j2\pi ({f_1} + {f_0} - {f_m})d}}} \right]\left[ {1 + {e^{j2\pi ({f_2} + {f_0})d}}} \right]} \\
&&\overline {\left[ {1 + {e^{ - j2\pi ({f_3} + {f_0})d}}} \right]\left[ {1 + {e^{j2\pi ({f_4} + {f_0} - {f_m})d}}} \right]} \\
&&\overline {{e^{j2\pi \left[ {({f_1} - {f_2})t + ({f_3} - {f_4})(t + \tau )} \right]}}d{f_1}d{f_2}d{f_3}d{f_4}}  \cdot {(\gamma /2)^2}\\
&=& \left[{2R({v_m})\!+\!R({v_m} \!+\! d){e^{j2\pi {f_0}d}}} { + R({v_m} \!-\! d){e^{ - j2\pi {f_0}d}}} \right]\!{e^{j\pi \!{f_m}\!{v_m}}}\\
&&\cdot\!\left[ {2{R^*}({v_m}) \!+\! {R^*}({v_m} + d){e^{ - j2\pi {f_0}d}}}{ + {R^*}({v_m} - d){e^{j2\pi {f_0}d}}} \right]\\
&&{e^{ - j\pi {f_m}{v_m}}}{e^{j2\pi {f_m}\tau }}{(\gamma /2)^2}.
\IEEEyesnumber
\end{IEEEeqnarray*}
Finally, the PSD of the signal can be obtained by the Fourier transform of $R_I^{sig}(\tau)$ \cite{Lu}, i.e.,
\begin{IEEEeqnarray*} {rCl}
\IEEEeqnarraymulticol{3}{l}{
{S^{sig}}(f)}\\
&=& \mathscr{F}\left[ {R_I^{sig}(\tau )} \right]\\
 &=& {(\gamma /2)^2}\left[ {{{\left| {H({v_m})} \right|}^2}\delta (f - {f_m}) + {{\left| {H( - {v_m})} \right|}^2}\delta (f + {f_m})} \right].\\
\IEEEyesnumber
\end{IEEEeqnarray*}
By integrating $S^{sig}(f)$ at $\pm f_m$, the power of the signal can be given by
\begin{equation}\label{eq:P_ssb_freq}
P_{ssb}^{sig} = {2(\gamma /2)^2}{\left| {H({v_m})} \right|^2}.
\end{equation}
We can see (\ref{eq:P_ssb_freq}) obtained by the frequency-domain analysis is as same as (\ref{eq:P_ssb}) obtained by the time-domain analysis.
\par On the other hand, the time averaged autocorrelation function of the noise $R_I^{no}(\tau)$ is given by (\ref{eq:R_I^no}) in the next page. As can be seem from (\ref{eq:R_I^no}), the autocorrelation function of the noise consists of 6 terms $r_1^{no}(\tau )$, $r_2^{no}(\tau )$, $r_3^{no}(\tau )$, $r_4^{no}(\tau )$, $r_5^{no}(\tau )$ and $r_6^{no}(\tau )$.

\begin{figure*}[!t]
\normalsize
\begin{IEEEeqnarray*} {rcl}
\IEEEeqnarraymulticol{3}{l}{
R_I^{no}(\tau )}\\
&=&\overline {\int\limits_{ - \infty }^{ + \infty } {\int\limits_{ - \infty }^{ + \infty } {\int\limits_{ - \infty }^{ + \infty } {\int\limits_{ - \infty }^{ + \infty } {\left\langle {{\tilde E_0}({f_1})\tilde E_0^*({f_4})} \right\rangle \left\langle {{\tilde E_0}({f_3})\tilde E_0^*({f_2})} \right\rangle } } } } } \\
&& \overline {\left[ {1 + {e^{ - j2\pi ({f_1} + {f_0})d}}} \right]\left[ {1 + {e^{j2\pi ({f_2} + {f_0})d}}} \right]\left[ {1 + {e^{ - j2\pi ({f_3} + {f_0})d}}} \right]\left[ {1 + {e^{j2\pi ({f_4} + {f_0})d}}} \right]{e^{j2\pi \left[ {({f_1} - {f_2})t + ({f_3} - {f_4})(t + \tau )} \right]}}d{f_1}d{f_2}d{f_3}d{f_4}}\\
 &+& {(\gamma /2)^2}\overline {\int\limits_{ - \infty }^{ + \infty } {\int\limits_{ - \infty }^{ + \infty } {\int\limits_{ - \infty }^{ + \infty } {\int\limits_{ - \infty }^{ + \infty } {\left\langle {{{\tilde E}_0}({f_1})\tilde E_0^*({f_4} - {f_m})} \right\rangle \left\langle {{{\tilde E}_0}({f_3} - {f_m})\tilde E_0^*({f_2})} \right\rangle } } } } } \\
&& \overline {\left[ {1\!+\! {e^{ - j2\pi ({f_1} + {f_0})d}}} \right]\!\left[ {1\!+\!{e^{j2\pi ({f_2} + {f_0})d}}} \right]\!\left[ {1\!+\!{e^{ - j2\pi ({f_3} + {f_0} - {f_m})d}}} \right]\!\left[ {1\!+\!{e^{j2\pi ({f_4} + {f_0} - {f_m})d}}} \right]\!{e^{j2\pi \left[ {({f_1} - {f_2})t + ({f_3} - {f_4})(t + \tau )} \right]}}d{f_1}d{f_2}d{f_3}d{f_4}}\\
 &+& {(\gamma /2)^2}\overline {\int\limits_{ - \infty }^{ + \infty } {\int\limits_{ - \infty }^{ + \infty } {\int\limits_{ - \infty }^{ + \infty } {\int\limits_{ - \infty }^{ + \infty } {\left\langle {{{\tilde E}_0}({f_3} - {f_m})\tilde E_0^*({f_2} - {f_m})} \right\rangle \left\langle {{{\tilde E}_0}({f_1})\tilde E_0^*({f_4})} \right\rangle } } } } } \\
&&\overline {\left[ {1 \!+\! {e^{ - j2\pi ({f_1} + {f_0})d}}} \right]\!\left[ {1 \!+\! {e^{j2\pi ({f_2} + {f_0} - {f_m})d}}}\! \right]\left[ {1 \!+\! {e^{ - j2\pi ({f_3} + {f_0} - {f_m})d}}} \right]\!\left[ {1 \!+\! {e^{j2\pi ({f_4} + {f_0})d}}} \right]\!{e^{j2\pi \left[ {({f_1} - {f_2})t + ({f_3} - {f_4})(t + \tau )} \right]}}d{f_1}d{f_2}d{f_3}d{f_4}}\\
 &+& {(\gamma /2)^2}\overline {\int\limits_{ - \infty }^{ + \infty } {\int\limits_{ - \infty }^{ + \infty } {\int\limits_{ - \infty }^{ + \infty } {\int\limits_{ - \infty }^{ + \infty } {\left\langle {{{\tilde E}_0}({f_3})\tilde E_0^*({f_2})} \right\rangle \left\langle {{{\tilde E}_0}({f_1} - {f_m})\tilde E_0^*({f_4} - {f_m})} \right\rangle } } } } } \\
&&\overline {\left[ {1 \!+\! {e^{ - j2\pi ({f_1} + {f_0} - {f_m})d}}} \right]\!\left[ {1 \!+\! {e^{j2\pi ({f_2} + {f_0})d}}} \right]\!\left[ {1 \!+\! {e^{ - j2\pi ({f_3} + {f_0})d}}} \right]\!\left[ {1 \!+\! {e^{j2\pi ({f_4} + {f_0} - {f_m})d}}} \right]\!{e^{j2\pi \left[ {({f_1} - {f_2})t + ({f_3} - {f_4})(t + \tau )} \right]}}d{f_1}d{f_2}d{f_3}d{f_4}}\\
 &+& {(\gamma /2)^2}\overline {\int\limits_{ - \infty }^{ + \infty } {\int\limits_{ - \infty }^{ + \infty } {\int\limits_{ - \infty }^{ + \infty } {\int\limits_{ - \infty }^{ + \infty } {\left\langle {{{\tilde E}_0}({f_3})\tilde E_0^*({f_2} - {f_m})} \right\rangle \left\langle {{{\tilde E}_0}({f_1} - {f_m})\tilde E_0^*({f_4})} \right\rangle } } } } } \\
&&\overline {\left[ {1 \!+\! {e^{ - j2\pi ({f_1} + {f_0} - {f_m})d}}} \right]\!\left[ {1 \!+\! {e^{j2\pi ({f_2} + {f_0} - {f_m})d}}} \right]\!\left[ {1 \!+\! {e^{ - j2\pi ({f_3} + {f_0})d}}} \right]\!\left[ {1 \!+\! {e^{j2\pi ({f_4} + {f_0})d}}} \right]\!{e^{j2\pi \left[ {({f_1} - {f_2})t + ({f_3} - {f_4})(t + \tau )} \right]}}d{f_1}d{f_2}d{f_3}d{f_4}}\\
 &+& {(\gamma /2)^4}\overline {\int\limits_{ - \infty }^{ + \infty } {\int\limits_{ - \infty }^{ + \infty } {\int\limits_{ - \infty }^{ + \infty } {\int\limits_{ - \infty }^{ + \infty } {\left\langle {{{\tilde E}_0}({f_3} - {f_m})\tilde E_0^*({f_2} - {f_m})} \right\rangle \left\langle {{{\tilde E}_0}({f_1} - {f_m})\tilde E_0^*({f_4} - {f_m})} \right\rangle } } } } } \\
&& \overline {\left[ {1 + {e^{ - j2\pi ({f_1} + {f_0} - {f_m})d}}} \right]\left[ {1 + {e^{j2\pi ({f_2} + {f_0} - {f_m})d}}} \right]\left[ {1 + {e^{ - j2\pi ({f_3} + {f_0} - {f_m})d}}} \right]\left[ {1 + {e^{j2\pi ({f_4} + {f_0} - {f_m})d}}} \right]} \\
 &&\cdot \overline {{e^{j2\pi \left[ {({f_1} - {f_2})t + ({f_3} - {f_4})(t + \tau )} \right]}}d{f_1}d{f_2}d{f_3}d{f_4}}\\
&=& r_1^{no}(\tau) + r_2^{no}(\tau) + r_3^{no}(\tau) + r_4^{no}(\tau) + r_4^{no}(\tau) + r_6^{no}(\tau)
\IEEEyesnumber \label{eq:R_I^no}
\end{IEEEeqnarray*}
\hrulefill
\vspace*{4pt}
\end{figure*}
The sum $r_1^{no}(\tau ) + r_3^{no}(\tau ) + r_4^{no}(\tau ) + r_6^{no}(\tau )$ selected from the 6 terms can be given by using (\ref{eq:xcor_freq}),
\begin{IEEEeqnarray*} {rcl}
\IEEEeqnarraymulticol{3}{l}{
r_1^{no}(\tau ) + r_3^{no}(\tau ) + r_4^{no}(\tau ) + r_6^{no}(\tau )}\\
 &=& \int\limits_{ - \infty }^{ + \infty } {G({f_2})\left[ {2 + 2\cos (2\pi ({f_2} + {f_0})d)} \right]}{\left| {T({f_2})} \right|^2}{e^{j2\pi {f_2}\tau }}d{f_2}\\
&& \int\limits_{ - \infty }^{ + \infty } {G({f_4})\left[ {2 + 2\cos (2\pi ({f_4} + {f_0})d)} \right]}{\left| {T({f_4})} \right|^2}{e^{ - j2\pi {f_4}\tau }}d{f_4}\\
&+& {(\gamma /2)^2}\int\limits_{ - \infty }^{ + \infty } {G({f_2} - {f_m})\left[ {2 + 2\cos (2\pi ({f_2} + {f_0} - {f_m})d)} \right]} \\
&& {\left| {T({f_2})} \right|^2}{e^{j2\pi {f_2}\tau }}d{f_2}\\
&& \int\limits_{ - \infty }^{ + \infty } {G({f_4})\left[ {2 + 2\cos (2\pi ({f_4} + {f_0})d)} \right]{\left| {T({f_4})} \right|^2}{e^{ - j2\pi {f_4}\tau }}d{f_4}} \\
&+& \int\limits_{ - \infty }^{ + \infty } {G({f_2})\left[ {2 + 2\cos (2\pi ({f_2} + {f_0})d)} \right]}{\left| {T({f_2})} \right|^2}{e^{j2\pi {f_2}\tau }}d{f_2} \\
&& \int\limits_{ - \infty }^{ + \infty } {G({f_4} - {f_m})\left[ {2 + 2\cos (2\pi ({f_4} + {f_0} - {f_m})d)} \right]} \\
&& {\left| {T({f_4})} \right|^2}{e^{ - j2\pi {f_4}\tau }}d{f_4} \cdot {(\gamma /2)^2}\\
&+& \int\limits_{ - \infty }^{ + \infty } {G({f_2} - {f_m})\left[ {2 + 2\cos (2\pi ({f_2} + {f_0} - {f_m})d)} \right]} \\
&& {\left| {T({f_2})} \right|^2}{e^{j2\pi {f_2}\tau }}d{f_2}\\
&& \int\limits_{ - \infty }^{ + \infty } {G({f_4} - {f_m})\left[ {2 + 2\cos (2\pi ({f_4} + {f_0} - {f_m})d)} \right]} \\
&& {\left| {T({f_4})} \right|^2}{e^{ - j2\pi {f_4}\tau }}d{f_4}{(\gamma /2)^4}.
\IEEEyesnumber
\end{IEEEeqnarray*}
Since $\int_{{\rm{ - }}\infty }^{{\rm{ + }}\infty } {G(f)\left[ {2 + 2\cos (2\pi (f + {f_0})d)} \right]{e^{j2\pi f\tau }}df}  = H(\tau )$, the sum $r_1^{no}(\tau ) + r_3^{no}(\tau ) + r_4^{no}(\tau ) + r_6^{no}(\tau )$ can be simplified to
\begin{IEEEeqnarray*} {rCl}
\IEEEeqnarraymulticol{3}{l}{
r_1^{no}(\tau ) + r_3^{no}(\tau ) + r_4^{no}(\tau ) + r_6^{no}(\tau )}\\
 = {\left| {H(\tau )} \right|^2}\left[ {1 + {{(\gamma /2)}^2}{e^{j2\pi {f_m}\tau }}} \right]\left[ {1 + {{(\gamma /2)}^2}{e^{ - j2\pi {f_m}\tau }}} \right].\\
\IEEEyesnumber
\end{IEEEeqnarray*}
Furthermore, $r_2^{no}(\tau )$ can be given by
\begin{IEEEeqnarray*} {rCl}
 r_2^{no}(\tau )&=& \int\limits_{ - \infty }^{ + \infty } {G({f_2})\left[ {2 + 2\cos (2\pi ({f_2} + {f_0})d)} \right]} \\
&&T({f_2} + {f_m}){T^*}({f_2}){e^{j2\pi {f_2}\tau }}d{f_2}\\
 &&\cdot\!\!\! \int\limits_{ - \infty }^{ + \infty } {G({f_4} - {f_m})\left[ {2 + 2\cos (2\pi ({f_4} + {f_0} - {f_m})d)} \right]} \\
&&T({f_4} - {f_m}){T^*}({f_4}){e^{ - j2\pi {f_4}\tau }}d{f_4} \cdot {e^{j2\pi {f_m}\tau }}{(\gamma /2)^2}\\
 &=& {e^{j2\pi {f_m}\tau }} \cdot H(\tau  - {v_m}){e^{ - j\pi {v_m}{f_m}}}\\
 &&\cdot {H^*}(\tau  - {v_m}){e^{ - j2\pi {f_m}(\tau  - {v_m})}}{e^{ - j\pi {v_m}{f_m}}} \cdot {(\gamma /2)^2}\\
 &=& {(\gamma /2)^2}{\left| {H(\tau  - {v_m})} \right|^2}.
\IEEEyesnumber
\end{IEEEeqnarray*}
Similarly, $r_5^{no}(\tau)$ can be given by
\begin{equation}
r_5^{no}(\tau ) = {(\gamma /2)^2}{\left| {H(\tau  + {v_m})} \right|^2}.
\end{equation}
Hence the overall autocorrelation function of the noise can be given by
\begin{IEEEeqnarray*} {rCl}
\IEEEeqnarraymulticol{3}{l}{
R_I^{no}(\tau )}\\
 &=& r_1^{no}(\tau ) + r_3^{no}(\tau ) + r_4^{no}(\tau ) + r_6^{no}(\tau ) + r_2^{no}(\tau ) + r_5^{no}(\tau )\\
 &=& {\left| {H(\tau )} \right|^2}\!\left[ {{\rm{1 + }}{{(\gamma /2)}^4} + {{(\gamma /2)}^2}{e^{j2\pi {f_m}\tau }} + {{(\gamma /2)}^2}{e^{ - j2\pi {f_m}\tau }}} \right]\\
 &&+ {\left| {H(\tau  - {v_m})} \right|^2}{(\gamma /2)^2} + {\left| {H(\tau  + {v_m})} \right|^2} \cdot {(\gamma /2)^2}.
\IEEEyesnumber
\end{IEEEeqnarray*}
Finally the PSD of the noise can be expressed as
\begin{IEEEeqnarray*} {rCl}
{S^{no}}(f) &=& \mathscr{F}\left[ {R_I^{no}(\tau )} \right]\\
 &=& {\left| {1 + {{(\gamma /2)}^2}{e^{j2\pi f{v_m}}}} \right|^2}{S_H}(f)\\
 &&+ {(\gamma /2)^2}{S_H}(f - {f_m}) + {(\gamma /2)^2}{S_H}(f + {f_m}).\\
 \IEEEyesnumber
\end{IEEEeqnarray*}
By evaluating the PSD of the noise at $\pm f_c$, the power of the noise is expressed as
\begin{IEEEeqnarray*} {rCl}
P_{ssb}^{no}({f_c}) &=& 2\left[ {1 + {{(\gamma /2)}^4} + 2{{(\gamma /2)}^2}\cos (2\pi {f_c}{v_c})} \right]{S_H}({f_c})\\
 &&+ 2{(\gamma /2)^2}{S_H}(2{f_c}) + 2{(\gamma /2)^2}{S_H}(0).
 \IEEEyesnumber \label{eq:P_no_freq}
\end{IEEEeqnarray*}
We can see (\ref{eq:P_no_freq}) is the same as (\ref{eq:Pssb_no}). By using (\ref{eq:P_ssb_freq}) and (\ref{eq:P_no_freq}), we can obtain the same SNR expression given in (\ref{eq:snr_ssb}).

\section*{Appendix II}
\begin{figure}[htbp]
\centering
\includegraphics [width=2.0in] {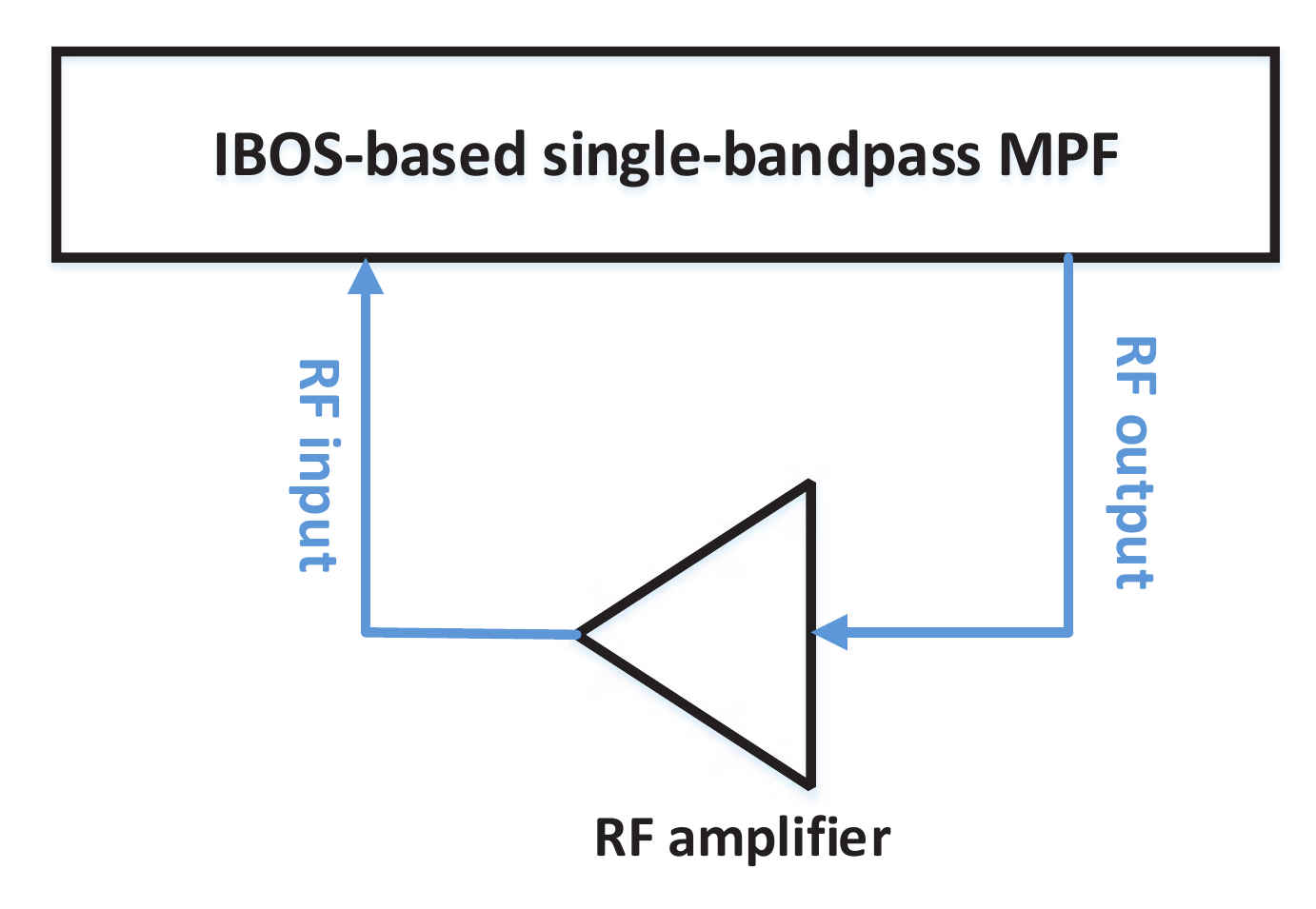}
\caption{Schematic diagram of OEO based on IBOS.}
\label{oeo}
\end{figure}
As demonstrated in \cite{Zhang}, OEO can be implemented by the IBOS-based single-bandpass MPF. The IBOS-based MPF acts as the mode-selecting device as well as the energy-storage element with high Q factor obtained by the low-loss fiber delay. The schematic diagram of the OEO is shown in Fig. \ref{oeo}. The RF amplifier is incorporated in the loop to compensate the loss of the MPF. When the OEO loop gain is greater than 0 dB, the OEO starts to oscillate. According to \cite{OEO}, the phase noise spectrum of generated RF signal can be expressed as
\begin{equation}\label{eq:phase noise}
{S_{RF}}(f') = \frac{\delta }{{2 - \delta /\tau  - 2\sqrt {1 - \delta /\tau } \cos (2\pi f'\tau )}}
\end{equation}
where $\delta$ is the input noise-to-signal ratio to the RF amplifier, $\tau$ is the total group delay of the OEO loop.
As can be seen from (\ref{eq:phase noise}), using the SNR of the IBOS-based MPF derived in (\ref{eq:snr_ssb}) or (\ref{eq:snr_pm}) and measuring the group delay of the OEO loop, the phase noise spectrum of the OEO can be obtained.

\end{document}